# Optical Second Harmonic Generation in Anisotropic Multilayers with Complete Multireflection of Linear and Nonlinear Waves using ♯SHAARP.*ml* Package

Rui Zu,[1,a] Bo Wang[1,5,a], Jingyang He[1], Lincoln Weber[1], Akash Saha[1], Long-Qing Chen[1,3,4], Venkatraman Gopalan[1,2,3]

[a]These authors contribute equally to this work

[1]Department of Materials Science and Engineering, The Pennsylvania State University, University Park, Pennsylvania 16802, USA

[2] Department of Physics, Pennsylvania State University, University Park, Pennsylvania, 16802, USA

[3]Department of Engineering Science and Mechanics, The Pennsylvania State University, University Park, Pennsylvania 16802, USA

[4]Department of Mathematics, The Pennsylvania State University, University Park, Pennsylvania 16802, USA

[5]Materials Science Division, Lawrence Livermore National Laboratory, Livermore, CA 94550, USA

Venkatraman Gopalan (vxg8@psu.edu); Long-Qing Chen (lqc3@psu.edu); Rui Zu (ruizu0110@gmail.com); Bo Wang (wang111@llnl.gov)

## Abstract

Optical second harmonic generation (SHG) is a nonlinear optical effect widely used for nonlinear optical microscopy and laser frequency conversion. Closed-form analytical solution of the nonlinear optical responses is essential for evaluating the optical responses of new materials whose optical properties are unknown *a priori*. A recent open-source code, ♯SHAARP.*si*, can provide such closed form solutions for crystals with arbitrary symmetries, orientations, and anisotropic properties at a single interface. However, optical components are often in the form of slabs, thin films on substrates, and multilayer heterostructures with multiple reflections of both the fundamental and up to ten different SHG waves at each interface, adding significant complexity. Many approximations have therefore been employed in the existing analytical approaches, such as slowly varying approximation, weak reflection of the nonlinear polarization, transparent medium, high crystallographic symmetry, Kleinman symmetry, easy crystal orientation along a high-symmetry direction, phase matching conditions and negligible interference among nonlinear waves, which may lead to large errors in the reported material properties. To avoid these approximations, we have developed an open-source package named Second Harmonic Analysis of Anisotropic Rotational Polarimetry in Multilayers (♯SHAARP.*ml*). The reliability and accuracy are established by experimentally benchmarking with both the SHG polarimetry and Maker fringes predicted from the package using reference materials as well as twisted 2-dimensional heterostructures.

GitHub: https://github.com/bzw133/SHAARP.ml

Documentation: https://shaarpml.readthedocs.io/en/latest/

## Introduction

The development of coherent laser light over a broad frequency spectrum from near-infrared and visible to terahertz (THz), ultraviolet, and X-rays regimes[1–4] has driven much of science and technology in the past decades, ranging from sensing, communications, biomedical instruments, imaging, and most recently nuclear fusion research.[5–10] Since the discovery of lasers in 1960 and the nonlinear optical effect in 1961[11,12], nonlinear optics has been a primary source for generating a continuously tunable electromagnetic spectrum. In the last two decades, quantum communications and computing have relied on using nonlinear optics to generate entangled photons and to achieve ultrafast all-optical switching.[13–15]

Optical second harmonic generation (SHG) refers to the nonlinear optical process where two photons of the same energy ($\hbar\omega$) combine to generate a new photon of higher energy ($2\hbar\omega$) in a nonlinear optical (NLO) medium. This phenomenon is described by the nonlinear polarization, $\mathbf{P}^{2\omega} = \boldsymbol{\chi}^{(2)}\mathbf{E}^{\omega}\mathbf{E}^{\omega}$, generated in the NLO material at $2\omega$ frequency by the electric field of the incident light, $\mathbf{E}^{\omega}$ at frequency $\omega$.[16] Here, $\boldsymbol{\chi}^{(2)}$ is the second-order nonlinear optical susceptibility represented by a third-rank tensor (with 18 independent components). If the refractive index and $\boldsymbol{\chi}^{(2)}$ tensors of a crystal are known, one can employ numerical simulations to model their nonlinear optical responses.[17,18] However, for new materials with unknown optical properties, experimental responses need to be measured and modeled by analytical or semi-analytical approaches to determine the coefficients by fitting the models to the experimental data. The complexity of developing such analytical models becomes untenable to perform manually when, in addition to the unknown $\boldsymbol{\chi}^{(2)}$ tensor, birefringence, arbitrary crystal symmetry and orientation, complex dielectric function, multilayer geometries, and interference of all the waves involved are considered. Large errors in $\boldsymbol{\chi}^{(2)}$ may be mistakenly introduced if the analysis is not handled properly.[19–21] The previous ♯SHAARP.*si* package addresses this need only for a single interface. The current ♯SHAARP.*ml* package addresses this need for realistic slabs and multilayer structures found in most optics applications.

**Table 1** summarizes the commonly applied models in existing SHG analyses. The foundation for the theoretical modeling of SHG responses was established by Maker, Bloembergen and Pershan (BP), Jerphagnon and Kurtz (JK), et al. in the 1960s and 1970s[22–24], for nonlinear optical processes in a transparent isotropic medium. In particular, the Maker fringe technique has become the primary method for characterizing nonlinear optical susceptibilities in transparent crystals, where the transmitted SHG intensity is measured as a function of the angle of incidence.[25–27] Further advances in the Maker fringes technique were made by Herman and Hayden (HH), and Shoji., et al., extending its applicability towards uniaxial systems and biaxial materials cut along high-symmetry directions.[28–30] However, these characterization methods are generally limited to transparent systems with high crystallographic symmetry, *p*- and *s*- polarized pump and SHG waves, and relatively simple geometry such as a bulk single crystal, a single-crystal slab, or a single-crystalline film on a substrate. SHG polarimetry is another technique to map out the anisotropic $\boldsymbol{\chi}^{(2)}$ by varying the fundamental and second harmonic polarization states of light, which is applicable to both transparent and absorbing crystals.[31–35] Nonetheless, the theoretical analyses for both Maker fringes and SHG polarimetry still involve many assumptions such as the slowly varying approximation, weak reflection of the nonlinear polarization, transparent medium, high crystallographic symmetry, Kleinman symmetry, easy crystal orientation along a high-symmetry direction, phase matching conditions ($n^{\omega} = n^{2\omega}$), and negligible interference among nonlinear waves.[23,24,29,36–40] Our existing package ♯SHAARP.*si* addressed arbitrary crystal symmetry, orientation, and complex dielectric function for a single interface.[21] However, its application requires analyzing nonlinear optical response in a single homogeneous crystal where the crystal is wedged to avoid specular reflections from the back surface (if the crystal is transparent), or the crystal has a thickness greater than the absorption depth for the fundamental and SHG waves (if the crystal is absorbing). To our best knowledge, there is no general tool available that can analytically or semi-analytically model, without the simplifying approximations made in BP, HH and JK models[23,24,28], the SHG responses of multilayer systems where light propagates through multiple layers of nonlinear optical materials, such as stacked 2D materials[41], near Fabry-Perot conditions[42], periodic domain gratings[43], and superlattices[44].

Table 1. Comparison of modeling capabilities among Bloembergen and Pershan Method (BP), Jerphagnon and Kurtz method (JK), Herman and Hayden method (HH), ♯SHAARP.*si,* and ♯SHAARP.*ml*.

| Features | BP | JK | HH | ♯SHAAR.*si* | ♯SHAAPR.*ml* |
|---|---|---|---|---|---|
| Probing geometry[a] | R and T | T | T | R | R and T |
| Layers[b] | SI or 1 | 1 | 2 | SI | Any |
| Symmetry | Isotropic | Isotropic | Uniaxial | Any | Any |
| Orientation[c] | × | High symmetry[a] | High symmetry[a] | Any | Any |
| Light polarization[d] | *p*- or *s*- | *p*- or *s*- | *p*- or *s*- | Any | Any |
| Absorption | √ | × | × | √ | √ |
| MR of $\mathbf{E}^{e\&o,\omega}$ [e] | × | × | × | N/A[f] | √ |
| MR of $\mathbf{E}^{e\&o,2\omega}$ [e] | √ | × | √ | N/A | √ |
| MR of $\mathbf{P}^{2\omega}$ [e] | × | × | × | N/A | √ |

[a] R and T refer to reflection and transmission, respectively.
[b] SI represents single interface. Numbers reflect the number of layers.
[c] High symmetry means samples are oriented along a high-symmetry direction.
[d] *p*- or *s*- refer to the electric fields of electromagnetic waves either parallel or perpendicular to the plane of incidence, respectively.
[e] MR represents multiple reflections of waves, $\mathbf{E}^{e\&o}$ represents homogeneous waves at their corresponding frequency, $\omega$ or $2\omega$ (e for extraordinary and o for ordinary waves), and $\mathbf{P}^{2\omega}$ stands for nonlinear polarization that gives rise to SHG effects.
[f] N/A refers to not applicable.

In this work, we present a comprehensive theoretical framework and an open-source package, ♯SHAARP.*ml* (Second Harmonic Analysis of Anisotropic Rotational Polarimetry for multilayers), for modeling second harmonic generation in an arbitrary single interface (same as ♯SHAARP.*si*)[21] and complex heterostructure with full consideration of multireflection at both linear and nonlinear frequencies. The ♯SHAARP.*ml* is designed to provide numerical and analytical nonlinear optical solutions for both simulation and experimental characterization, allowing for fast, flexible, and user-friendly analysis of nonlinear optical response on complex material systems. Five key attributes of ♯SHAARP.*ml* include: (1) ability to model a multilayer stack with an arbitrary number of layers with homogeneous optical properties, (2) allowing arbitrary crystallographic symmetry, orientation, and possess absorption, birefringence, and dispersion of each layer, (3) choices for both reflection and transmission probing geometries, (4) full control of the polarization states of the incident and detected waves, and (5) explicit consideration of the multireflection of both linear and the nonlinear waves. While there are other contributions to the SHG response, such as magnetically induced SHG[45], electric quadrupole[46], etc., we focus on the electric dipole SHG in this work and shall extend the ♯SHAARP package to include other sources of SHG in future studies.

Seven materials systems were used to benchmark the analysis using the ♯SHAARP.*ml* package: α-quartz single crystal, Au-coated α-quartz bilayer, $LiNbO_3$ and KTP single crystals, ZnO//Pt//$Al_2O_3$ thin film and multiple SHG active layers ($LiNbO_3$//quartz, and twisted bilayer $MoS_2$). Good agreement between results from ♯SHAARP.*ml* and the literature on the measured SHG coefficients for standard single crystal materials demonstrate the accuracy and reliability of the package.

## Results and Discussion

### Theoretical background

**Figure 1a** presents the ray diagram of linear and nonlinear waves through a multilayer system adopted in ♯SHAARP.*ml*. Without loss of generality, we assume the first layer ($M_1$) to be SHG active. In a more general case, all layers can (but need not) be SHG active in experiments. When a monochromatic plane wave at $\omega$ frequency is incident upon the system, the electromagnetic properties of the plane wave inside the system are governed by the wave equation at $\omega$ frequency,

$$\nabla \times \nabla \times \mathbf{E}^{\omega} + \begin{pmatrix} \tilde{\varepsilon}^{\omega}_{L_1L_1} & \tilde{\varepsilon}^{\omega}_{L_1L_2} & \tilde{\varepsilon}^{\omega}_{L_1L_3} \\ \tilde{\varepsilon}^{\omega}_{L_2L_1} & \tilde{\varepsilon}^{\omega}_{L_2L_2} & \tilde{\varepsilon}^{\omega}_{L_2L_3} \\ \tilde{\varepsilon}^{\omega}_{L_3L_1} & \tilde{\varepsilon}^{\omega}_{L_3L_2} & \tilde{\varepsilon}^{\omega}_{L_3L_3} \end{pmatrix} \boldsymbol{\mu}^{\omega} \frac{\partial^2}{\partial t^2} \mathbf{E}^{\omega} = 0 \tag{1}$$

where $\boldsymbol{E}^{\omega}$, $\tilde{\varepsilon}^{\omega}_{L_iL_j}$ and $\boldsymbol{\mu}^{\omega}$ are respectively the electric field inside the medium at $\omega$ frequency, anisotropic dielectric tensor components in the lab coordinate system (LCS), and magnetic permeability tensor at $\omega$ frequency. The $\boldsymbol{\mu}^{\omega}$ will be assumed to be vacuum permeability for a nonmagnetic system, $\boldsymbol{\mu}^{\omega} \sim \mu_0 \boldsymbol{I}$, where $\boldsymbol{I}$ is the identity matrix. The subscripts $i$ and $j$ are dummy indices describing the direction of each tensor component of the anisotropic dielectric susceptibility tensor in the LCS, denoted as $\tilde{\boldsymbol{\varepsilon}}^{\omega}_{\mathrm{LCS}}$. Note that $\tilde{\boldsymbol{\varepsilon}}^{\omega}_{\mathrm{LCS}}$ can be complex to account for absorption. Four coordinate systems are utilized, namely, principal coordinate system (PCS), crystal physics coordinate system (ZCS), crystallographic coordinate system (CCS), and lab coordinate system (LCS). In PCS, the complex dielectric susceptibility tensor is diagonalized. ZCS is the orthogonal coordinate system in which the property tensors are defined, such as dielectric susceptibility tensor, SHG tensor, piezoelectricity tensor, etc.[47] The CCS describes the coordinate system formed by the basis vectors of the unit cell (which are not necessarily orthogonal), and LCS is an orthogonal coordinate system of the model system with the plane of incidence (PoI) coincides with the $L_1$-$L_3$ plane as shown in **Figure 1a**. Note that PCS, ZCS, and LCS are orthogonal coordinate systems, while the CCS can be non-orthogonal depending on the crystal symmetry. **Equation (1)** is a generalized eigenvalue problem that can be solved routinely.[48] The resulting eigenvalues and eigenvectors are related to the effective refractive indices and electric field directions for both ordinary and extraordinary waves. Due to reflectance at various interfaces, both forward and backward propagating waves exist in the heterostructure. The resulting backward propagating wavevectors can be described as

$$(\mathbf{k}^{\mathrm{eB},\omega}, \mathbf{k}^{\mathrm{oB},\omega})_{\mathrm{M}_i} = \left( \begin{pmatrix} 1 & 0 & 0 \\ 0 & 1 & 0 \\ 0 & 0 & -1 \end{pmatrix} \cdot \mathbf{k}^{\mathrm{eF},\omega}, \begin{pmatrix} 1 & 0 & 0 \\ 0 & 1 & 0 \\ 0 & 0 & -1 \end{pmatrix} \cdot \mathbf{k}^{\mathrm{oF},\omega} \right)_{\mathrm{M}_i}, 1 \le i \le N \tag{2}$$

The superscripts e, o, F, and B, respectively, represent extraordinary, ordinary, forward-propagating, and backward-propagating waves. $\mathrm{M}_i$ represents the $i^{\mathrm{th}}$ medium in the heterostructure. Similarly, the full electromagnetic properties of backward propagating waves can be obtained using **Equations (1)** and **(2)**. The boundary conditions require the tangential components of both wave vectors and field strengths to be continuous across the interface where the former relation yields Snell's law, and the latter represents the Fresnel coefficients. Thus the propagation direction, effective refractive indices, and field strengths can be obtained by simultaneously solving the equations below,[23,49]

$$k_{L_1}^{\mathrm{i},\omega} = k_{L_1}^{\mathrm{R},\omega} = \left(k_{L_1}^{\mathrm{eF},\omega}\right)_{\mathrm{M}_i} = \left(k_{L_1}^{\mathrm{oF},\omega}\right)_{\mathrm{M}_i} = \left(k_{L_1}^{\mathrm{eB},\omega}\right)_{\mathrm{M}_i} = \left(k_{L_1}^{\mathrm{oB},\omega}\right)_{\mathrm{M}_i} = k_{L_1}^{\mathrm{T},\omega}, 1 \le i \le N \tag{3}$$

$$E_{\parallel}^{\mathrm{i},\omega} + E_{\parallel}^{\mathrm{R},\omega} = \left(E_{\parallel}^{\mathrm{eF},\omega} + E_{\parallel}^{\mathrm{oF},\omega} + E_{\parallel}^{\mathrm{eB},\omega} + E_{\parallel}^{\mathrm{oB},\omega}\right)_{\mathrm{M}_1} \tag{4}$$

$$\left(E_{\parallel}^{\mathrm{eF},\omega} e^{i\phi^{\mathrm{eF},\omega}} + E_{\parallel}^{\mathrm{oF},\omega} e^{i\phi^{\mathrm{oF},\omega}} + E_{\parallel}^{\mathrm{eB},\omega} e^{-i\phi^{\mathrm{eB},\omega}} + E_{\parallel}^{\mathrm{oB},\omega} e^{-i\phi^{\mathrm{oB},\omega}}\right)_{\mathrm{M}_i} = \left(E_{\parallel}^{\mathrm{eF},\omega} + E_{\parallel}^{\mathrm{oF},\omega} + E_{\parallel}^{\mathrm{eB},\omega} + E_{\parallel}^{\mathrm{oB},\omega}\right)_{\mathrm{M}_{i+1}}, 1 \le i \le N-1 \tag{5}$$

$$\left(E_{\parallel}^{\mathrm{eF},\omega} e^{i\phi^{\mathrm{eF},\omega}} + E_{\parallel}^{\mathrm{oF},\omega} e^{i\phi^{\mathrm{oF},\omega}} + E_{\parallel}^{\mathrm{eB},\omega} e^{-i\phi^{\mathrm{eB},\omega}} + E_{\parallel}^{\mathrm{oB},\omega} e^{-i\phi^{\mathrm{oB},\omega}}\right)_{\mathrm{M}_N} = E_{\parallel}^{\mathrm{T},\omega} \tag{6}$$

$$H_{\parallel}^{\mathrm{i},\omega} + H_{\parallel}^{\mathrm{R},\omega} = \left(H_{\parallel}^{\mathrm{eF},\omega} + H_{\parallel}^{\mathrm{oF},\omega} + H_{\parallel}^{\mathrm{eB},\omega} + H_{\parallel}^{\mathrm{oB},\omega}\right)_{\mathrm{M}_1} \tag{7}$$

$$\left(H_{\parallel}^{\mathrm{eF},\omega} e^{i\phi^{\mathrm{eF},\omega}} + H_{\parallel}^{\mathrm{oF},\omega} e^{i\phi^{\mathrm{oF},\omega}} + H_{\parallel}^{\mathrm{eB},\omega} e^{-i\phi^{\mathrm{eB},\omega}} + H_{\parallel}^{\mathrm{oB},\omega} e^{-i\phi^{\mathrm{oB},\omega}}\right)_{\mathrm{M}_i} = \left(H_{\parallel}^{\mathrm{eF},\omega} + H_{\parallel}^{\mathrm{oF},\omega} + H_{\parallel}^{\mathrm{eB},\omega} + H_{\parallel}^{\mathrm{oB},\omega}\right)_{\mathrm{M}_{i+1}}, 1 \le i \le N-1 \tag{8}$$

$$\left(H_{\parallel}^{\mathrm{eF},\omega} e^{i\phi^{\mathrm{eF},\omega}} + H_{\parallel}^{\mathrm{oF},\omega} e^{i\phi^{\mathrm{oF},\omega}} + H_{\parallel}^{\mathrm{eB},\omega} e^{-i\phi^{\mathrm{eB},\omega}} + H_{\parallel}^{\mathrm{oB},\omega} e^{-i\phi^{\mathrm{oB},\omega}}\right)_{\mathrm{M}_N} = E_{\parallel}^{\mathrm{T},\omega} \tag{9}$$

Here, $\phi$ is the phase difference for a forward wave propagating from top to bottom surfaces and for a backward wave propagating from bottom to top surfaces of layer M$i$, defined as $\phi = h_{\mathrm{M}_i}\mathbf{k} \cdot (0,0,-1)$, where $h_{\mathrm{M}_i}$ is the thickness of the $i^{\mathrm{th}}$ medium. The subscript ‖ indicates tangential components along both $L_1$ and $L_2$ directions. **Equations (3)** – **(9)** can be expanded depending on the number of layers in the heterostructure.

The optical dipolar second harmonic generation is defined by the generation of nonlinear polarization at $2\omega$ frequency when the NLO materials are pumped by the incident electric fields at $\omega$ frequency. The nonlinear polarization is defined as

$$\mathbf{P}_{\mathrm{M}_i}^{2\omega} = \varepsilon_0 \boldsymbol{\chi}^{(2)} \mathbf{E}_{\mathrm{M}_i}^{\omega} \mathbf{E}_{\mathrm{M}_i}^{\omega} e^{i(\mathbf{k}^{\mathrm{S}} \cdot \mathbf{r} - 2\omega t)} \tag{10}$$

where $\mathbf{P}_{\mathrm{M}_i}^{2\omega}$, $\mathbf{E}_{\mathrm{M}_i}^{\omega}$, $\varepsilon_0$, $\boldsymbol{\chi}^{(2)}$, $\mathbf{k}^{\mathrm{S}}$ and $\mathbf{r}$ are nonlinear polarization, fundamental electric field, vacuum dielectric permittivity, second-order nonlinear optical susceptibility, wave vector of the source wave, and position vector, respectively. Since arbitrary layers can be SHG active, $\mathbf{P}_{\mathrm{M}_i}^{2\omega}$ will appear when the $i^{\mathrm{th}}$ layer is SHG active, as denoted by the subscript $\mathrm{M}_i$. The generated nonlinear polarization is often known as the source wave that gives rise to the nonlinear optical effects. It is important to note that during the propagation of fundamental fields, the nonlinear polarization is generated throughout the entire optical path of $\mathbf{E}_{\mathrm{M}_i}^{\omega}$, according to **equation (10)**. When the multiple reflections of nonlinear polarization are considered, the interference of nonlinear polarization can be obtained by considering the multiple reflections of $\mathbf{E}_{\mathrm{M}_i}^{\omega}$. Many previous theoretical studies of transmission SHG assume weak reflection of the source wave and ignore the multi-reflection of nonlinear polarization[24,28]. Though a few other works considered the multiple reflections explicitly[30,50,51], they rely on approximations such as high symmetry structures with high symmetry axes aligned along the probing directions.

The propagation of nonlinear waves is governed by the wave equation at $2\omega$ frequency, written as

$$\nabla \times \nabla \times \mathbf{E}^{2\omega} + \begin{pmatrix} \tilde{\varepsilon}_{L_1L_1}^{2\omega} & \tilde{\varepsilon}_{L_1L_2}^{2\omega} & \tilde{\varepsilon}_{L_1L_3}^{2\omega} \\ \tilde{\varepsilon}_{L_2L_1}^{2\omega} & \tilde{\varepsilon}_{L_2L_2}^{2\omega} & \tilde{\varepsilon}_{L_2L_3}^{2\omega} \\ \tilde{\varepsilon}_{L_3L_1}^{2\omega} & \tilde{\varepsilon}_{L_3L_2}^{2\omega} & \tilde{\varepsilon}_{L_3L_3}^{2\omega} \end{pmatrix} \boldsymbol{\mu}^{2\omega} \frac{\partial^2}{\partial t^2} \mathbf{E}^{2\omega} = -\boldsymbol{\mu}^{2\omega} \frac{\partial^2}{\partial t^2} \mathbf{P}^{2\omega} \tag{11}$$

where $\mathbf{P}^{2\omega}$, $\mathbf{E}^{2\omega}$, $\tilde{\varepsilon}_{L_iL_j}^{2\omega}$, and $\boldsymbol{\mu}^{2\omega}$ are nonlinear polarization, radiated electric field, the component of complex dielectric permittivity tensor in LCS ($\tilde{\boldsymbol{\varepsilon}}_{\mathrm{LCS}}^{2\omega}$), and magnetic permeability tensor at $2\omega$ frequency. **Equation (11)** highlights the fundamental mechanisms of nonlinear optics, where the generated nonlinear polarization works as a source wave, generating and radiating second harmonic electric fields that can freely propagate inside the medium. Therefore, the particular and general solutions of **equation (11)** correspond to the bound and free waves, respectively.[28] The propagation of $\mathbf{P}^{2\omega}$ is confined to the propagation of the fundamental wave at $\omega$ that generates it, and the corresponding $\mathbf{E}^{2\omega}$ is hence called the bound wave or inhomogeneous wave. On the other hand, the SHG wave generated by the bound wave can freely propagate governed by the direction specified by Snell's law at $2\omega$, hence it is called the free wave or the homogeneous wave.

The anisotropic three-wave mixing phenomena is revealed in **equation (10)**, where material anisotropy is taken into account. In each SHG active medium (M$i$), the forward and backward nonlinear wavevectors can thus be identified as $\mathbf{k}^{\mathrm{S},2\omega} = 2\mathbf{k}^{\mathrm{eF},\omega}$, $2\mathbf{k}^{\mathrm{oF},\omega}$, $\mathbf{k}^{\mathrm{eF},\omega} + \mathbf{k}^{\mathrm{oF},\omega}$, $2\mathbf{k}^{\mathrm{eB},\omega}$, $2\mathbf{k}^{\mathrm{oB},\omega}$, $\mathbf{k}^{\mathrm{eB},\omega} + \mathbf{k}^{\mathrm{oB},\omega}$, $\mathbf{k}^{\mathrm{eF},\omega} + \mathbf{k}^{\mathrm{eB},\omega}$, $\mathbf{k}^{\mathrm{eF},\omega} + \mathbf{k}^{\mathrm{oB},\omega}$, $\mathbf{k}^{\mathrm{oF},\omega} + \mathbf{k}^{\mathrm{eB},\omega}$, and $\mathbf{k}^{\mathrm{oF},\omega} + \mathbf{k}^{\mathrm{oB},\omega}$. The wavevectors for the ten nonlinear polarizations in the $i^{\mathrm{th}}$ layer are thus denoted as $(\mathbf{k}^{\mathrm{eFeF},2\omega}$, $\mathbf{k}^{\mathrm{oFoF},2\omega}$, $\mathbf{k}^{\mathrm{eFoF},2\omega}$, $\mathbf{k}^{\mathrm{eBeB},2\omega}$, $\mathbf{k}^{\mathrm{oBoB},2\omega}$, $\mathbf{k}^{\mathrm{eBoB},2\omega}$, $\mathbf{k}^{\mathrm{eFeB},2\omega}$, $\mathbf{k}^{\mathrm{eFoB},2\omega}$, $\mathbf{k}^{\mathrm{oFeB},2\omega}$, $\mathbf{k}^{\mathrm{oFoB},2\omega})_{\mathrm{M}i}$ for clarity, as shown in **Figure 1a**. For example, a nonlinear polarization $\mathbf{P}^{\mathrm{eFoB},2\omega}$ is formed when a forward propagating extraordinary wave ($\mathbf{k}^{\mathrm{eF},\omega}$) and a backward propagating ordinary wave ($\mathbf{k}^{\mathrm{oB},\omega}$) are combined. However, the wave mixing terms containing both forward and backward waves, such as $\mathbf{k}^{\mathrm{eFeB},2\omega}$ and $\mathbf{k}^{\mathrm{oFoB},2\omega}$, are often dropped or ignored in existing literature due to a large phase mismatch.[29,30] Although these terms form standing waves propagating parallel to the layers, the standing waves at both the top and bottom surfaces of each layer can still contribute to the boundary conditions. For example, a nonlinear polarization ($\mathbf{P}^{\mathrm{eFeB},2\omega}$) can be generated by a mixture of $\mathbf{k}^{\mathrm{eF},\omega}$ and $\mathbf{k}^{\mathrm{eB},\omega}$ at top or bottom surfaces leading to additional components in the boundary conditions. Therefore, we have implemented the mixing term in ♯SHAARP.*ml*, resulting in, at most, ten distinct nonlinear polarizations of different combinations of wavevectors for each SHG active layer. These ten waves are shown as ten different arrows in **Fig. 1a**.

The particular solutions of **equation (11)** can be obtained using the method described in previous work.[21] For example, the electric field of the nonlinear polarization induced by the mixture of two forward extraordinary waves can be written as $\mathbf{E}^{\mathrm{eFeF},2\omega} = \mathbf{C}^{\mathrm{eFeF},2\omega} e^{i(\mathbf{k}^{\mathrm{eFeF},2\omega}\cdot\mathbf{r}-2\omega t)}$, where $\mathbf{C}^{\mathrm{eFeF},2\omega}$ is a vector describing the direction and magnitude of the resulting bounded electric field due to the nonlinear polarization. Thus, all electric and magnetic fields generated by the ten distinct nonlinear polarizations can be uniquely identified by solving **equation (11)**. On the other hand, the general solution of **equation (11)**, which represents the homogeneous waves, can be calculated following the same procedure as solving **equation (1)** but at $2\omega$ frequency. Four nonlinear waves will be obtained to fully describe the multiple reflections of homogeneous waves, namely, $(\mathbf{E}^{\mathrm{eF},2\omega}, \mathbf{E}^{\mathrm{oF},2\omega}, \mathbf{E}^{\mathrm{eB},2\omega}, \mathbf{E}^{\mathrm{oB},2\omega})_{\mathrm{M}i}$, whose field strengths are determined using the boundary conditions to be described below.

The momentum conservation and energy conservation of the generated $2\omega$ waves lead to the following boundary condition:

$$k_{L_1}^{\mathrm{R},2\omega} = \left(k_{L_1}^{\mathrm{eF},2\omega}\right)_{\mathrm{M}_i} = \left(k_{L_1}^{\mathrm{oF},2\omega}\right)_{\mathrm{M}_i} = \left(k_{L_1}^{\mathrm{eB},2\omega}\right)_{\mathrm{M}_i} = \left(k_{L_1}^{\mathrm{oB},2\omega}\right)_{\mathrm{M}_i} = k_{L_1}^{\mathrm{T},2\omega}, 1 \le i \le N \quad (12)$$

$$E_{\parallel}^{\mathrm{R},2\omega} = (E_{\parallel}^{\mathrm{eF},2\omega} + E_{\parallel}^{\mathrm{oF},2\omega} + E_{\parallel}^{\mathrm{eB},2\omega} + E_{\parallel}^{\mathrm{oB},2\omega} + E_{\parallel}^{\mathrm{eFeF},2\omega} + E_{\parallel}^{\mathrm{oFoF},2\omega} + E_{\parallel}^{\mathrm{eFoF},2\omega} + E_{\parallel}^{\mathrm{eBeB},2\omega} + E_{\parallel}^{\mathrm{oBoB},2\omega} + E_{\parallel}^{\mathrm{eBoB},2\omega} + E_{\parallel}^{\mathrm{eFeB},2\omega} + E_{\parallel}^{\mathrm{eFoB},2\omega} + E_{\parallel}^{\mathrm{oFeB},2\omega} + E_{\parallel}^{\mathrm{oFoB},2\omega})_{\mathrm{M}_1} \quad (13)$$

$$(E_{\parallel}^{\mathrm{eF},2\omega} e^{i\phi^{\mathrm{eF},2\omega}} + E_{\parallel}^{\mathrm{oF},2\omega} e^{i\phi^{\mathrm{oF},2\omega}} + E_{\parallel}^{\mathrm{eB},2\omega} e^{-i\phi^{\mathrm{eB},2\omega}} + E_{\parallel}^{\mathrm{oB},2\omega} e^{-i\phi^{\mathrm{oB},2\omega}} + E_{\parallel}^{\mathrm{eFeF},2\omega} e^{i(2\phi^{\mathrm{eF},\omega})} + E_{\parallel}^{\mathrm{oFoF},2\omega} e^{i(2\phi^{\mathrm{oF},\omega})} + E_{\parallel}^{\mathrm{eFoF},2\omega} e^{i(\phi^{\mathrm{eF},\omega}+\phi^{\mathrm{oF},\omega})} + E_{\parallel}^{\mathrm{eBeB},2\omega} e^{-i(2\phi^{\mathrm{eB},\omega})} + E_{\parallel}^{\mathrm{oBoB},2\omega} e^{-i(2\phi^{\mathrm{oB},\omega})} + E_{\parallel}^{\mathrm{eFeB},2\omega} + E_{\parallel}^{\mathrm{eFoB},2\omega} e^{i(\phi^{\mathrm{eF},\omega}-\phi^{\mathrm{oB},\omega})} + E_{\parallel}^{\mathrm{oFeB},2\omega} e^{i(\phi^{\mathrm{oF},\omega}-\phi^{\mathrm{eB},\omega})} + E_{\parallel}^{\mathrm{oFoB},2\omega})_{\mathrm{M}_i} = (E_{\parallel}^{\mathrm{eF},2\omega} + E_{\parallel}^{\mathrm{oF},2\omega} + E_{\parallel}^{\mathrm{eB},2\omega} + E_{\parallel}^{\mathrm{oB},2\omega} + E_{\parallel}^{\mathrm{eFeF},2\omega} + E_{\parallel}^{\mathrm{oFoF},2\omega} + E_{\parallel}^{\mathrm{eFoF},2\omega} + E_{\parallel}^{\mathrm{eBeB},2\omega} + E_{\parallel}^{\mathrm{oBoB},2\omega} + + E_{\parallel}^{\mathrm{eBoB},2\omega} + E_{\parallel}^{\mathrm{eFeB},2\omega} + E_{\parallel}^{\mathrm{eFoB},2\omega} + E_{\parallel}^{\mathrm{oFeB},2\omega} + E_{\parallel}^{\mathrm{oFoB},2\omega})_{\mathrm{M}_{i+1}}, 1 \le i \le N-1 \quad (14)$$

$$(E_{\parallel}^{\mathrm{eF},2\omega} + E_{\parallel}^{\mathrm{oF},2\omega} + E_{\parallel}^{\mathrm{eB},2\omega} + E_{\parallel}^{\mathrm{oB},2\omega} + E_{\parallel}^{\mathrm{eFeF},2\omega} + E_{\parallel}^{\mathrm{oFoF},2\omega} + E_{\parallel}^{\mathrm{eFoF},2\omega} + E_{\parallel}^{\mathrm{eBeB},2\omega} + E_{\parallel}^{\mathrm{oBoB},2\omega} + E_{\parallel}^{\mathrm{eBoB},2\omega} + E_{\parallel}^{\mathrm{eFeB},2\omega} + E_{\parallel}^{\mathrm{eFoB},2\omega} + E_{\parallel}^{\mathrm{oFeB},2\omega} + E_{\parallel}^{\mathrm{oFoB},2\omega})_{\mathrm{M}_N} = E_{\parallel}^{\mathrm{T},2\omega} \quad (15)$$

$$H_{\parallel}^{\mathrm{R},2\omega} = (H_{\parallel}^{\mathrm{eF},2\omega} + H_{\parallel}^{\mathrm{oF},2\omega} + H_{\parallel}^{\mathrm{eB},2\omega} + H_{\parallel}^{\mathrm{oB},2\omega} + H_{\parallel}^{\mathrm{eFeF},2\omega} + H_{\parallel}^{\mathrm{oFoF},2\omega} + H_{\parallel}^{\mathrm{eFoF},2\omega} + H_{\parallel}^{\mathrm{eBeB},2\omega} + H_{\parallel}^{\mathrm{oBoB},2\omega} + H_{\parallel}^{\mathrm{eBoB},2\omega} + H_{\parallel}^{\mathrm{eFeB},2\omega} + H_{\parallel}^{\mathrm{eFoB},2\omega} + H_{\parallel}^{\mathrm{oFeB},2\omega} + H_{\parallel}^{\mathrm{oFoB},2\omega})_{\mathrm{M}_1} \quad (16)$$

$$(H_{\parallel}^{\mathrm{eF},2\omega} e^{i\phi^{\mathrm{eF},2\omega}} + H_{\parallel}^{\mathrm{oF},2\omega} e^{i\phi^{\mathrm{oF},2\omega}} + H_{\parallel}^{\mathrm{eB},2\omega} e^{-i\phi^{\mathrm{eB},2\omega}} + H_{\parallel}^{\mathrm{oB},2\omega} e^{-i\phi^{\mathrm{oB},2\omega}} + H_{\parallel}^{\mathrm{eFeF},2\omega} e^{i(2\phi^{\mathrm{eF},\omega})} + H_{\parallel}^{\mathrm{oFoF},2\omega} e^{i(2\phi^{\mathrm{oF},\omega})} + H_{\parallel}^{\mathrm{eFoF},2\omega} e^{i(\phi^{\mathrm{eF},\omega}+\phi^{\mathrm{oF},\omega})} + H_{\parallel}^{\mathrm{eBeB},2\omega} e^{-i(2\phi^{\mathrm{eB},\omega})} + H_{\parallel}^{\mathrm{oBoB},2\omega} e^{-i(2\phi^{\mathrm{oB},\omega})} + H_{\parallel}^{\mathrm{eFeB},2\omega} + H_{\parallel}^{\mathrm{eFoB},2\omega} e^{i(\phi^{\mathrm{eF},\omega}-\phi^{\mathrm{oB},\omega})} + H_{\parallel}^{\mathrm{oFeB},2\omega} e^{i(\phi^{\mathrm{oF},\omega}-\phi^{\mathrm{eB},\omega})} + H_{\parallel}^{\mathrm{oFoB},2\omega})_{\mathrm{M}_i} = (H_{\parallel}^{\mathrm{eF},2\omega} + H_{\parallel}^{\mathrm{oF},2\omega} + H_{\parallel}^{\mathrm{eB},2\omega} + H_{\parallel}^{\mathrm{oB},2\omega} + H_{\parallel}^{\mathrm{eFeF},2\omega} + H_{\parallel}^{\mathrm{oFoF},2\omega} + H_{\parallel}^{\mathrm{eFoF},2\omega} + H_{\parallel}^{\mathrm{eBeB},2\omega} + H_{\parallel}^{\mathrm{oBoB},2\omega} + + H_{\parallel}^{\mathrm{eBoB},2\omega} + H_{\parallel}^{\mathrm{eFeB},2\omega} + H_{\parallel}^{\mathrm{eFoB},2\omega} + H_{\parallel}^{\mathrm{oFeB},2\omega} + H_{\parallel}^{\mathrm{oFoB},2\omega})_{\mathrm{M}_{i+1}}, 1 \leq i \leq N-1 \quad (17)$$

$$(H_{\parallel}^{\mathrm{eF},2\omega} + H_{\parallel}^{\mathrm{oF},2\omega} + H_{\parallel}^{\mathrm{eB},2\omega} + H_{\parallel}^{\mathrm{oB},2\omega} + H_{\parallel}^{\mathrm{eFeF},2\omega} + H_{\parallel}^{\mathrm{oFoF},2\omega} + H_{\parallel}^{\mathrm{eFoF},2\omega} + H_{\parallel}^{\mathrm{eBeB},2\omega} + H_{\parallel}^{\mathrm{oBoB},2\omega} + H_{\parallel}^{\mathrm{eBoB},2\omega} + H_{\parallel}^{\mathrm{eFeB},2\omega} + H_{\parallel}^{\mathrm{eFoB},2\omega} + H_{\parallel}^{\mathrm{oFeB},2\omega} + H_{\parallel}^{\mathrm{oFoB},2\omega})_{\mathrm{M}_N} = H_{\parallel}^{\mathrm{T},2\omega} \quad (18)$$

where $\phi$ is the phase difference for a forward wave propagating from top to bottom surface and for a backward wave propagating from bottom to top surface in layer $\mathrm{M}_i$, defined as $\phi = h_{\mathrm{M}_i}\mathbf{k} \cdot (0,0,-1)$. **Equations (12)-(18)** describe the most general case where all layers are SHG active, except for the air layers. For a non-SHG active layer, all the fields of the inhomogeneous waves will be zero due to the absence of nonlinear polarization while the homogeneous $2\omega$ waves will still be present. For a standing wave formed at either the top or bottom surface in the medium $\mathrm{M}_i$, taking $E_{\parallel}^{\mathrm{eFeB},2\omega}$ as an example, the phase terms are mutually canceled out, leading to the same field strength at both interfaces. Finally, with all the nonlinear waves and boundary conditions considered, both polarization-resolved reflected and transmitted SHG intensities can be obtained.

The SHG measurement geometry is shown in **Figure 1b**, where the incident light (red) is focused on the surface of the sample (a heterostructure labeled by $\mathrm{M}_1$ to $\mathrm{M}_N$), and the generated SHG response can be collected in either transmission or reflection geometry. With this measurement geometry, two common techniques, namely SHG polarimetry and Maker fringe methods, can be deployed to probe the SHG tensors of nonlinear optical materials. For SHG polarimetry measurement, both the incident polarization ($\varphi$, polarizer) and SHG polarization ($\psi$, analyzer) can be varied to probe the polarization-dependent anisotropic SHG tensor. This method provides more comprehensive information on the anisotropy than the Maker fringe method and can be utilized to identify the orientation and the point group symmetry of a crystal. On the other hand, the Maker fringes method measures the transmitted SHG response as a function of angle of incidence ($\theta^{\mathrm{i}}$) with fixed polarization directions of both the incident and the SHG waves, such as *p*- or *s*- polarized light waves. The variation in the envelope of the SHG intensity versus the angle of incidence can reveal the relative magnitude of nonlinear susceptibilities. However, the transmission geometry for Maker fringes limits its applications to material systems that are transparent. ♯SHAARP.*ml* can model both the SHG polarimetry and Maker fringes numerically or semi-analytically, which can be used to determine the unknown SHG tensors of new materials.

**Outline of ♯SHAARP**

The theoretical method described in the preceding section is implemented using Wolfram Mathematica with a user-friendly GUI and a detailed tutorial, which can be found in Ref.[52]. Following the naming convention of our previous work, we named the newly developed software capable of modeling optical SHG of multilayer system as ♯SHAARP.*ml*. **Figure 2** illustrates the calculation procedure of ♯SHAARP.*ml*. New features compared with ♯SHAARP.*si* are the capability of handling more than one interface, the addition of backward propagating waves and resulting nonlinear polarizations, phase accumulation of light, etc. First, with a given point group symmetry, the dielectric tensor in the ZCS, and its orientation relative to the LCS coordinate system as inputs, one can conveniently obtain the mutual relations among the four coordinate systems within ♯SHAARP.*ml*, and thus define the geometry of the system. Then, by solving the wave equation with the boundary conditions at $\omega$ frequencies, one can obtain the forward and backward propagating waves in each layer, $(\mathbf{E}^{\mathrm{eF},\omega}, \mathbf{E}^{\mathrm{oF},\omega}, \mathbf{E}^{\mathrm{eB},\omega}, \mathbf{E}^{\mathrm{oB},\omega})_{\mathrm{M}_i}$. The obtained sets of field strengths are the result of multiple reflections at the pump frequency.[49] The generated nonlinear polarization vectors can thus be obtained from electric fields at $\omega$ frequency. Further solving the wave equation at the nonlinear frequency can provide the wavevector and electric field directions of all forward and backward homogeneous and inhomogeneous waves in each layer (14 waves in each NLO layer, 4 waves in the non-NLO layer). Finally, plugging all the waves at $2\omega$ frequency into the boundary conditions of electric and magnetic fields gives transmitted and reflected polarization-resolved nonlinear optical response.

**Case studies using ♯SHAARP.*ml***

In the following, we present our experimental measurements of the SHG responses for a few typical nonlinear optical crystals and their heterostructures to demonstrate how they can be interpreted by numerical and semi-analytical analyses using ♯SHAARP.*ml*. In particular, we studied the Maker fringes of pure and Au-coated quartz single crystals and the SHG polarimetry of $LiNbO_3$, KTP, and $ZnO//Pt//Al_2O_3$ heterostructure. We also performed two predictive modelings of bilayers consisting of two SHG active materials, namely, $LiNbO_3//\alpha\text{-}SiO_2$ and twisted bilayer $MoS_2$, which can be helpful in distinguishing ferroelectric domain states and nonlinear optical studies in low dimensional material systems. These examples not only serve as benchmark tests of ♯SHAARP.*ml* against known NLO materials covering a wide range of types (uniaxial, biaxial, and absorbing) but also demonstrate the broad applicability of ♯SHAARP.*ml* to a variety of situations (e.g., Maker fringes, polarimetry, quantifying the effect of adopting different assumptions in the SHG modeling, analytical fitting to extract absolute values of SHG coefficients, and predictive simulations of SHG responses of NLO heterostructures).

**Maker fringes of α-quartz single crystal**

The study of α-quartz in nonlinear optics can be traced back to the discovery of second harmonic generation in 1961.[11] The first benchmark study for ♯SHAARP.*ml* is performed using the single crystalline α-quartz, which has been extensively investigated previously using the Maker fringes method.[22,24,28,30] The SHG coefficient $d_{11}$ has been measured to be $0.3\ \mathrm{pm/V}$.[53] In this case study, we demonstrate the capability of ♯SHAARP.*ml* in obtaining the semi-analytical expression for Maker fringe response and benchmark analysis with both existing models in the literature[24,28] and our experimental investigations. **Figure 3** shows the comparison among numerical simulation results from ♯SHAARP.*ml* with various modeling conditions and existing results using analytical methods.[24,28] The Maker fringes condition is summarized in **Figure 3a.** The fundamental wavelength ($\lambda^{\omega}$) is 1064 nm and the generated SHG signal from a 300 μm X-cut quartz is analyzed. Both the fundamental and SHG waves are *p*- polarized. Two widely applied Maker fringes models are utilized for comparison, namely the JK (Jerphagnon & Kurtz[24]) method and HH (Herman & Hayden[28]) method. The JK method was developed for an isotropic medium with an assumption that only forward propagating waves are involved.[24] The HH method extended this model to a birefringent uniaxial system with multiple reflections of homogeneous waves (free waves) at $2\omega$ frequency, but not for the inhomogeneous waves or linear waves. ♯SHAARP.*ml* involves multiple reflections for both linear and nonlinear waves (homogeneous and inhomogeneous) and thus can be reduced to JK or HH methods by making the corresponding assumptions. Schematics of the assumptions made for the three approaches can be found in **Supplementary Note 1**, **Figure S1**. **Figures 3b** and **3c** illustrate the three Maker fringes patterns obtained from the HH method (denoted as analytic HH) and numerical analysis using ♯SHAARP.*ml* with both JK and HH modeling conditions, denoted as ♯SHAARP(JK) and ♯SHAARP(HH).[24,28] The blue dots, yellow and green lines correspond to analytic HH,

♯SHAARP(JK) and ♯SHAARP(HH), respectively. All three Maker fringe patterns are consistent with the literature.[28] In particular, analytic HH and ♯SHAARP(HH) show good agreement, demonstrating ♯SHAARP.*ml* can accurately reproduce the prior results. **Figure 3c** shows the magnified area of the dashed box region in **Figure 3b**. By enabling the multiple reflections of homogeneous waves at $2\omega$ frequency, ♯SHAARP(HH) produce additional fine fringes at $\theta^i$ from 20° to 30°, which are absent for ♯SHAARP(JK). This difference indicates that the interference between forward and backward homogenous $2\omega$ waves results in these fine fringes.

To demonstrate the effect of full multiple reflection (FMR) in determining the nonlinear optical responses, we performed a comparative study to measure the Maker fringes of uncoated and Au-coated quartz slabs, as shown in **Figure 4**. **Figure 4a** and **4b** show the experimental conditions and corresponding Maker fringes patterns using a 123.6 μm uncoated Z-cut quartz slab. The incident fundamental wave is *p*- polarized centered at 800 nm, and the generated *p*-polarized SHG intensity is collected as a function of $\theta^i$. Four Maker fringes patterns are compared, namely, experimental results (Expt.), ♯SHAARP(JK), ♯SHAARP(HH), and full multiple reflections of linear and nonlinear waves (♯SHAARP(FMR)). Due to the weak reflectance of quartz, all three modeling conditions yield similar Maker fringes patterns, in agreement with the experimental results. The centers of the fringes overlap with that of ♯SHAARP(JK). The major difference lies in the fine fringes of the Maker fringes patterns, as highlighted in the inset of **Figure 4b** (a zoom-in of the dashed regions near $\theta^i = 30°$). With more multiple reflections considered and thus more interferences, the amplitude of the fine fringes increases. Experimentally, the fine fringes are not observable with a fine step size of $\theta^i$ at 0.1°, and possible reasons for not detecting fine fringes can be the range of angles of incidence, the nonuniformity of sample thickness within the probing area, or the bandwidth of the laser. To confirm the above effects, Maker fringes patterns with averaging angle of incidence (due to beam divergence of ~3°), thickness variation (of ~50nm across the beam), and wavelength averaging ($\lambda^{\omega} \pm 5$ nm) are performed (see **Supplementary Note 2**, **Figure S2a** and **S2b**). It is found that by averaging the above three parameters one can effectively smoothen the calculated Maker fringes pattern, confirming that the variation of experimental conditions, as used for the case of quartz, can smear out the fine fringes. Averaging $\theta^{i}$ and $\lambda^{\omega}$ have a more dominating effect compared with averaging $h$ for the case of quartz in this study. It is important to note that although the JK method can also produce a smooth Maker fringes pattern, this coincidence is accidental. In fact, the smooth pattern obtained by averaging the angle of incidence correctly considers the multiple reflection of waves and the variation of experimental conditions while the JK method excludes the fine fringes due to the neglect of reflective waves.

To illustrate the circumstance under which FMR becomes critical, we further studied the Maker fringes of a Z-cut quartz with Au coating at the backside of the slab, as shown in **Figures 4c** and **4d**. The thickness and complex refractive index of Au coating are determined by spectroscopic ellipsometry (see **Supplementary Note 3**, **Figure S3**). The thickness of the Au layer is found to be 13.9 nm, far below the penetration depth (~45 nm). Due to the strong reflection of the Au layer, the resulting backward propagating waves are expected to be more intense than those in the pure quartz case. To test such hypothesis, we compared the simulation results based on ♯SHAARP(HH) and ♯SHAARP(FMR) against the experimental results, as shown in **Figure 4d**. Due to the inclusion of Au, the fine fringes resulting from multiple reflections become more prominent as compared with **Figure 4b**. Similar phenomenon has also been observed in other studies.[30,50,54] It can be seen from **Figure 4d** that ♯SHAARP(HH) fails to capture the total transmitted SHG intensity and the relative intensity ratio between $\theta^{i} = 0°$ and $\approx$ 40°. In contrast, the results from ♯SHAARP(FMR) indicate better agreement with experiments regarding these SHG intensities but exhibit large variation in the fine fringes that are smeared out in the experiments. These oscillations can be corrected by averaging the angle of incidence, thickness variation in the probed area, and finite bandwidth of the fundamental wavelength, leading to the results denoted as ♯SHAARP(FMR+$\theta^i$+$h$+$\lambda^\omega$). Detailed discussion on the corrections can be found in **Supplementary Note 2**, **Figure S2c** and **S2d**. With ♯SHAARP(FMR+$\theta^i$+$h$+$\lambda^\omega$), the SHG relative intensities, peak, and minimum positions are well captured simultaneously with good agreement between the experiments.

In contrast to the fine fringes originating from the interference of the fundamental waves, the broader envelope in the SHG intensity with respect to $\theta^i$ (interval ranging across tens of degrees visible in **Figures 3b**, **4b** and **4d**) carry the essential information associated with the interference between the homogeneous and inhomogeneous waves. This interference originates from the phase difference between the source waves ($\mathbf{k}^{\mathrm{S},2\omega}$) and the homogeneous waves ($\mathbf{k}^{\mathrm{e},2\omega}$ and $\mathbf{k}^{\mathrm{o},2\omega}$) accumulated throughout the bilayer structure, and thus, the broader envelope is extremely sensitive to the changes in the crystal thickness and refractive indices at both $\omega$ and $2\omega$ frequencies. Therefore, SHG Maker fringes can be utilized as a sensitive probe of wafer uniformity.[55] For example, with a thickness variation of 1 $\mu m$, the Maker fringes change drastically, as demonstrated in **Supplementary Note 4** (see **Figure S4**). It is worth noting that the crystal thicknesses determined in **Figures 4b** and **4d** are slightly different, i.e., 123.6 $\mu m$ and 121.4 $\mu m$, respectively, due to the change of probing positions and nonuniform thickness across the sample (10 μm variation across a 10 mm × 10 mm sample), as confirmed by the stylus profilometry. In addition, we note that the example presented in **Figure 4c** and **d** also illustrates the capability of ♯SHAARP.*ml* in handling multiple layers with strong reflections.

The phase difference between two propagating waves is critical to determining their interference, e.g., being constructive and destructive for in-phase and out-of-phase situations respectively. With ♯SHAARP.*ml*, we show that different ways to compute the relative phase terms of the waves can lead to dissimilar results. Conventionally, the phases of electromagnetic waves propagating through layers are calculated as $\phi = h_{\mathrm{M}i}\mathbf{k} \cdot (0,0,-1)$, where only the $L_3$ component of the wavevector is considered. On the other hand, the full phase of the electromagnetic wave accumulated through layers can be written as $\phi = h_{\mathrm{M}i}\mathbf{k} \cdot (\tan\theta, 0, -1)$, where $\theta$ represents the refractive angle of the corresponding wave. However, the Maker fringes obtained using full phase show large deviation from the experiments (see **Supplementary Note 5**, **Figure S5**). Such discrepancy may come from the fact that a small beam size comparative to the crystal thickness is used in the experiment, where a sizeable beam overlap and finite resolution of angles are essential for the interference to become observable in the experiments. Therefore, for the quartz case, taking only the vertical phase along $L_3$ direction will be sufficient in the SHG analysis throughout the current work.

**$LiNbO_3$ and KTP Single Crystals**

$LiNbO_3$ and $KTiOPO_4$ (potassium titanyl phosphate, KTP) have been widely studied for decades owing to their excellent nonlinear optical properties.[56–58] Their well-established nonlinear optical susceptibilities make the two crystals suitable for benchmarking analysis. Utilizing the partial analytical expressions generated by ♯SHAARP.*ml*, the experimental polarimetry results can be analyzed to extract relative ratios of SHG coefficients, and the absolute SHG coefficients of the two single crystals can be obtained using α-quartz as the reference.

$LiNbO_3$ crystallizes in a trigonal structure with the point group $3m$ and has a bandgap of around 3.8 eV.[57,59] Two orientations, namely (0001) (i.e., Z-cut) and $(11\bar{2}0)$ (i.e., X-cut) were measured in the transmission geometry and analyzed simultaneously to determine the full SHG tensor using a fundamental wavelength ($\lambda^{\omega}$) center at 1550 nm. **Figures 5a-5d** show the experimental results and fitting analysis of $LiNbO_3$. Three angles of incidence ($\theta^i = 0°, 10°, 30°$) are analyzed simultaneously, and the SHG intensities are normalized within each orientation. **Figures 5a** and **5b** are the SHG polarimetry results of ~538 µm thick $LiNbO_3$ $(11\bar{2}0)$ crystal slab, whose *c* axis is placed along the $L_1$ direction (see the experimental orientations in **Supplementary Note 6**, **Figure S6**). The obtained polar plots are *p*- and *s*- polarized SHG intensities as a function of incident polarization ($\varphi$). The dominating $d_{33}$ (corresponds to $\theta^i = 0$ in **Figure 5a**) results in a large intensity difference between the *p*- and *s*- polarized SHG responses (~135 times difference), which can be well captured by ♯SHAARP.*ml*. **Figures 5c** and **5d** are the measured SHG intensities and fitting results of ~119 µm $LiNbO_3$ (0001). At normal incidence, both *p*- and *s*- polarized SHG show four lobes with equal intensities arising from the in-plane isotropy in this orientation. As the crystal is tilted towards a larger angle of incidence, the projection of $d_{33}$ to the $L_1$ increases, leading to an increase in the *p*- polarized SHG intensity, as seen in **Figure 5c**. By fitting two $LiNbO_3$ crystals with different orientations and using quartz as the reference, the extracted ratios and absolute values of the SHG coefficients of $LiNbO_3$ are summarized in **Table 2**, which agree well with previously reported values[24,29].

KTP adopts an orthorhombic crystal structure with a point group of *mm2*. It is classified as a biaxial material with distinct optical responses along all three crystal physics axes. Thus, a careful analysis of full anisotropy and the presence of two optical axes are critical in optical modeling. In this study, we used two KTP slabs simultaneously, namely ~370 µm X-cut ((100) orientation) slab and ~570 μm Y-cut ((010) orientation) slab, to analyze the full SHG tensor. Both $c$ axes are placed along the $L_2$ direction, and their two optical axes lie in $Z_1$-$Z_3$ plane (see the experimental orientations in **Supplementary Note 6**, **Figure S6**).[47,60] **Figures 5e** and **5f** are the SHG polar plots for $p$- and $s$- polarized SHG response, respectively. Four angles of incidence are utilized to identify five unknown SHG susceptibilities uniquely ($\theta^{\mathrm{i}} = 0°, 10°, 20°$,and $40°$). Using partial analytical expressions generated by ♯SHAARP.*ml*, the SHG polarimetry fittings show good agreement between the theory and experimental data, and the extracted ratios and absolute values of SHG coefficients of KTP are summarized in **Table 2**.

As discussed in previous work,[21] the symmetry assumptions, such as that of isotropy, can lead to errors of up to 30% in the ratios between SHG coefficients, depending on the anisotropy of the materials. In this work, our discussion will focus on the influence of Kleinman's symmetry (KS), the exclusion of multireflection of linear waves and nonlinear inhomogeneous waves (NMR), and the exclusion of the nonlinear polarizations formed by the mixture of forward and backward waves (No $\mathbf{P}^{\mathrm{FB},2\omega}$). Using ♯SHAARP, these three factors can be selectively applied in the modeling and fitting analysis to investigate the influence of individual assumptions on the final obtained nonlinear susceptibilities. **Figures 5g** and **5h** summarize the SHG coefficients ratios obtained under different assumptions to fit the same experimental data for $LiNbO_3$ and KTP, respectively. The Kleinman's symmetry (KS) assumes all three indices in the $d$ tensor are permutable, leading to $d_{31}$= $d_{15}$ in $LiNbO_3$, and $d_{31}$= $d_{15}$, $d_{32}$= $d_{24}$ in KTP.[61–65] The NMR case is equivalent to the HH method, where only multiple reflections of the nonlinear homogeneous wave are considered. The "No $\mathbf{P}^{\mathrm{FB},2\omega}$" case neglects the nonlinear polarizations generated by mixed forward and backward waves, i.e., $\mathbf{P}^{\mathrm{eFeB},2\omega}$, $\mathbf{P}^{\mathrm{eFoB},2\omega}$, $\mathbf{P}^{\mathrm{oFeB},2\omega}$, and $\mathbf{P}^{\mathrm{oFoB},2\omega}$. The ♯SHAARP.*ml* case represents the analysis with full consideration of multireflection of linear and nonlinear waves, all possible nonlinear polarizations and complete material anisotropy, and no Kleinmann symmetry assumed. Comparing the four cases, we found most of the obtained SHG ratios vary within 20-30%, which are commonly comparable to the error bars. The NMR case is close to the ♯SHAARP.*ml* case, implying that the HH method may be a good approximation for studying KTP with photon energies below its bandgap. The difference between "♯SHAARP.*ml*" and "No $\mathbf{P}^{\mathrm{FB.2}\omega}$" lies in whether or not to include the nonlinear polarization created by the interference of forward and backward electric fields at $\omega$ frequency. For transparent crystals (such as $LiNbO_3$ and KTP) whose reflectance is smaller, the backward propagating electric fields are weaker, and thus the resulting $\mathbf{P}^{\mathrm{FB,2\omega}}$ does not contribute much to the final transmitted SHG intensities. This observation also supports the claims of early models that only nonlinear polarization formed by forward propagating waves is considered for transparent samples.[23,24,28] The KS case, however, can introduce relatively large deviations in the obtained coefficient ratios such as a 60% error for $d_{31}/d_{32}$ in KTP.

**ZnO//Pt//$Al_2O_3$ thin films**

ZnO has been widely studied for decades for electronics, photonics, and optoelectronics applications owing to its large piezoelectric coefficients, large exciton binding energies, wide optical bandgap, and good chemical and thermal stability.[66–68] Recently, ZnO with Mg substitution ($Zn_{1-x}Mg_xO$) has been shown to possess ferroelectricity, paving its way toward waveguides and quasi-phase-matched (QPM) frequency conversion devices.[15,43] Though the nonlinear optical process in ZnO has been extensively explored in both bulk and thin films forms, its nonlinear optical susceptibilities have been reported with a large scatter in the values from less than one pm/V to hundreds of pm/V, indicating either sample variations or inconsistent modeling of the SHG data.[69–72] In this work, we select 159 nm ZnO//200nm Pt//0.5mm $Al_2O_3$ as an example to demonstrate the capabilities of ♯SHAARP.*ml* in probing thin films on substrates with a bottom electrode and the importance of multiple reflections in the analysis.

As described in earlier work, ZnO was grown using RF magnetron sputtering and formed a stack of ZnO//Pt//$Al_2O_3$, as shown in **Figure 6a**.[43] The fundamental wavelength is centered at 1550 nm, and the angle of incidence is set to 45 degrees ($\theta^i = 45°$). The reflected $p$- and $s$-polarized SHG intensities at 775 nm are then collected as a function of incident polarization (azimuthal angle $\varphi$). The epitaxial ZnO (0001) films adopt the wurtzite structure (point group *6mm*) and remain isotropic within the in-plane direction. **Figure 6b** shows the crystal structure of ZnO and its crystallographic directions relative to the lab coordinate systems, where $Z_1 \parallel L_1$ and $Z_3 \parallel L_3$. Due to the strong reflection of the Pt bottom electrode at fundamental 1550 nm and SHG wavelength at 775 nm, the multiple reflections at both frequencies inside the ZnO layer are thus significant and cannot be ignored. The thickness of the Pt layer is around 200 nm, and therefore both incident and SHG waves will be fully blocked and reflected by the Pt layer. Since earlier theoretical approaches often assume weak reflection of the nonlinear source wave,[23,24,28] a suitable theoretical model in the literature that can resolve a near Fabry-Perot condition is difficult to find. Using expressions generated by ♯SHAARP.*ml*, the experimental results can thus be fitted, as demonstrated in **Figures 6c** and **6d**.

Further reference against a wedged X-cut $LiNbO_3$ yields the absolute SHG coefficients of the entire SHG tensor. **Figure 6e** summarizes the absolute SHG coefficients obtained from ♯SHAARP.*ml* in comparison with the cases under various assumptions (the meanings of the notations are consistent with the previous section). The "♯SHAARP.*ml*" case yields the absolute $d_{33} = 6.6 \pm 2.2$ pm/V, which is close to early reported values for films and single crystals (~7.15 pm/V).[70] This indicates the film under study has good qualities and low optical loss.

Comparing the results from ♯SHAARP.*ml* with those from KS and No $\mathbf{P}^{\mathrm{FB}}$, the obtained absolute SHG coefficients are reasonably close. On the other hand, the multiple reflections play a more significant role in the analysis. As can be seen from the NMR case, the obtained nonlinear susceptibilities are greatly exaggerated by one order of magnitude. This is because the total SHG signals were attributed to the single propagation of nonlinear polarization from the top to the bottom surface instead of multiple bounces. To compensate for the path difference between NMR and FMR, the nonlinear susceptibilities have to be increased, leading to $d^{\mathrm{SHG}}$ of nearly 10 times higher than the actual value.

This case study of ZnO thin films highlights the necessity of a more general nonlinear optical model because of the increased complexity as more materials are involved in a heterostructure. For example, SHG has been widely applied in characterizing 2D materials on top of $SiO_2$//Si substrate which is highly reflective in the visible regime.[32,73] Nevertheless, the multireflection of the heterostructure is often assumed to be negligible.[74] Additionally, as more binary ferroelectric semiconductors are being discovered, such as (Al,Sc)N and (Al,B)N,[42,75,76] optical second harmonic generation as a non-destructive method will be a unique tool for probing ferroelectricity. The ZnO//Pt//$Al_2O_3$ case shown here highlights the capability of ♯SHAARP.*ml* not only in handling various probing geometries (transmission and reflection) that goes beyond the well-established Maker fringes method but also in modeling heterostructures near the Fabry-Perot condition. In particular, the analytical and numerical approaches enabled by ♯SHAARP.*ml* provide versatile solutions for the purpose of materials characterization and numerical simulation.

**Table 2** summarizes the absolute nonlinear optical susceptibilities and their relative ratios of all four crystalline materials obtained from this work and reported in literature. The accuracy of ♯SHAARP.*ml* is benchmarked, covering single crystals and thin film-based heterostructure, material systems that are highly transparent or reflective, and distinct anisotropy from uniaxial to biaxial optical classes.

Table 2. Comparison of absolute SHG coefficients and their ratios between ♯SHAARP and literature. Absolute values are in the unit of pm/V. The SHG coefficients of quartz and $MoS_2$ used in this work are adopted from the literature.

| Materials | Wavelenth, $\lambda$ (nm) | SHG Coefficients | This work | Ref[27,32,53,57] |
|---|---|---|---|---|
| $LiNbO_3$ slab | 1550 | $\|d_{33}\|$ | $19.3 \pm 0.6$ | $18.9 \pm 2.1$ |
| | | $d_{33}/d_{31}$ | $5.5 \pm 0.5$ | $6.1 \pm 0.7$ |
| | | $d_{22}/d_{31}$ | $-0.4 \pm 0.1$ | $-0.5 \pm 0.1$ |
| KTP slab | 1550 | $\|d_{33}\|$ | $12.8 \pm 0.1$ | $12.6 \pm 0.6$ |
| | | $d_{33}/d_{32}$ | $4.2 \pm 0.1$ | $3.9 \pm 0.4$ |
| | | $d_{31}/d_{32}$ | $0.8 \pm 0.3$ | $0.6 \pm 0.1$ |
| | | $d_{24}/d_{32}$ | $1.0 \pm 0.1$ | $1.0 \pm 0.1$ |
| | | $d_{15}/d_{32}$ | $0.4 \pm 0.1$ | $0.5 \pm 0.1$ |
| $ZnO//Pt//Al_2O_3$ | 1550 | $\|d_{33}\|$ | $\mp 6.6 \pm 2.2$ | $-7.2$ |
| | | $\|d_{31}\|$ | $\pm 0.8 \pm 0.3$ | 0.7 |
| | | $\|d_{15}\|$ | $\pm 1.1 \pm 0.1$ | 1 (KS) |
| α-quartz (w/wo Au) | 800, 1550 | $\|d_{11}\|$ | 0.3[a] | 0.3[a] |
| | | $\|d_{14}\|$ | 0 | 0 |
| $MoS_2$ Bilayer | 810 | $\|d_{22}\|$ | 158 | 158 |

[a]The values are converted to corresponding wavelength using the Miller's rule before comparison.[87]

**SHG active bilayers, $LiNbO_3$//Quartz**

The generated SHG signals, in general, contain both amplitude and phase information of materials, such as the direction of a static (zero frequency) spontaneous polarization, $P_s$, of ferroelectric materials. (Note that this static ferroelectric polarization is distinct from any optical polarization at optical frequencies we have discussed earlier). Two ferroelectric domains with antiparallel spontaneous polarizations (separated by a 180° domain wall) will generate nonlinear optical polarizations with a $\pi$ phase shift, yet of the same amplitude. Thus, the corresponding

SHG intensities are identical for the two domains, leaving the ferroelectric domain state indistinguishable based on the intensity alone.[16,33,77] The SHG interference contrast imaging has been developed to resolve this issue.[45,77–80] In this subsection, we employ ♯SHAARP.*ml* simulation to illustrate the basic idea of SHG interference contrast imaging, intimately (without an air gap in this example) placing a periodically poled $LiNbO_3$ $(2\bar{1}\bar{1}0)$ crystal on top of a Z-cut quartz crystal as a model system.

The principle of SHG interference contrast imaging is schematically shown in **Figure 7a**, where the blue and red are fundamental waves and SHG waves, respectively.[16,45] An additional quartz is placed beneath the $LiNbO_3$ crystal (abbreviated as LNO) to generate the interference of the nonlinear waves through reflection. The nonlinear waves generated by $LiNbO_3$ (denoted as $\mathbf{k}_{\mathrm{L}}^{2\omega}$) and quartz (denoted as $\mathbf{k}_{\mathrm{Q}}^{2\omega}$) will interfere to resolve the phase information of $\mathbf{k}_{\mathrm{L}}^{2\omega}$. **Figure 7b** shows four cases where Case 1 and 4 involve only $LiNbO_3$ $(2\bar{1}\bar{1}0)$ crystals with opposite polarization directions and Case 2 and 3 have an identical (001) quartz layer placed under the $LiNbO_3$. The thicknesses of both $LiNbO_3$ and quartz are assumed to be 50 μm and 35 μm, respectively, and the fundamental light is set at normal incidence ($\theta^i = 0°$) with a wavelength centered at 1550 nm ($\lambda^\omega = 1550$ nm). **Figure 7c** and **7d** show simulation results of the SHG responses for the four cases using ♯SHAARP.*ml*. The simulated SHG polarimetry responses ($I_{L_1}^{2\omega}(\varphi)$ and $I_{L_2}^{2\omega}(\varphi)$) as a function of the incident optical polarization ($\varphi$) is illustrated in **Figure 7c**. The pure LNO cases with opposite ferroelectric polarization directions (cases 1 and 4) show identical SHG responses that cannot be distinguished from SHG polarimetry. In contrast, by placing the quartz below the LNO, the corresponding SHG responses between Cases 2 and 3 show a clear change. We pick $I_{L_1}^{2\omega}(0)$ for comparison among the four cases (**Figure 7d**) since when $\varphi = 0$, the light polarization at $\omega$ and $2\omega$ are parallel to the ferroelectric polarization, $P_s$, of $LiNbO_3$, giving rise to the largest SHG intensity. The intensities of the SHG waves in Cases 1 and 4 are the same while they are different in Cases 2 and 3. This is because the nonlinear waves generated by LNO ($\mathbf{k}_{\mathrm{L}}^{2\omega}$) and quartz ($\mathbf{k}_{\mathrm{Q}}^{2\omega}$) interfere constructively in Case 2 and destructively in Case 3. Thereby, the two ferroelectric domain states of $LiNbO_3$ can be differentiated by measuring the SHG intensity with the aid of a quartz reference layer. Beyond this example, ♯SHAARP.*ml* can easily handle extending this problem to include many SHG active layers and with arbitrary direction of ferroelectric polarization as long as each layer is homogeneous.

**Twisted bilayer $MoS_2$**

Nonlinear optical probes have been widely applied in the studies of two-dimensional material systems due to their sensitivity to structure, orientation, electronic structure, and material compositions. Twisted bilayer $MoS_2$ is one of the examples that contains two SHG active layers that are offset by certain degrees. The SHG signal from these bilayer systems is often modeled as an interference effect between the second harmonic waves arising from each monolayer with a phase difference depending on the twist angle.[79,81] **Figure 8** shows the ♯SHAARP.*ml* simulation results of twisted bilayer $MoS_2$ for various twist angles ($\Delta\theta$) in the rotating polarizer, rotating analyzer configuration (RA) at normal incidence as described in reference.[79] The material orientations and stacking orders are shown in **Figure 8(a)**. The bottom blue layer has a fixed orientation where the b-axis (zig-zag) is placed along the $L_2$ direction (i.e., $MoS_2$($\theta = 0°$)). The top orange layer ($MoS_2$($\theta$)) is then rotated counterclockwise by $\Delta\theta$. The complete stacking order is thus $MoS_2$($\theta = \Delta\theta$)//$MoS_2$($\theta = 0°$)//$Al_2O_3$(0001). In this case study, the fundamental wavelenghth is set at 800 nm, where $MoS_2$ remain transparent at $\omega$ frequency but highly resonating at $2\omega$ frequency with established refractive indices.[82,83] **Figure 8(b)** illustrates the SHG intensity variations and effective orientation change of monolayer $MoS_2$ with $\theta = 0°$ or 25° and the bilayer $MoS_2(\theta = 25°)//MoS_2(\theta = 0°)$. The reflected light with parallel polarizer and analyzer condition is chosen, i.e., $I_{\parallel}^{R,2\omega}$. For monolayers, the peaks of the lobes in **Figure 8(b)** are perpendicular to *a*, *b* and *a*+*b* directions. On the other hand, the peaks of bilayers are located between the peaks of monolayers, consistent with the experimental observations.[79,84] With a thickness of $Al_2O_3$ set at 501.93 um, the observed intensity ratio between the bilayer and monolayer is around 2.1, close to the intensity ratio observed experimentally.[79] It is worth noting that this ratio varies periodically between ~2.1 to ~2.4 with varying $Al_2O_3$ thickness (a few hundred micrometers). To further explore the FMR effects induced by the substrate, we use $\kappa(\Delta\theta) \equiv \frac{I_B - 2I_M}{2I_M}$ as the indicator, where $I_B$ and $I_M$ are peak intensities of bilayer and monolayer, as depicted in **Figure 8(c)**. In a simplified situation where only the anisotropic SHG tensor is considered, it can be derived that $\kappa(\Delta\theta) = \cos 3\Delta\theta$, as indicated by the black curve.[79] When $\Delta\theta = 0°$, the two monolayers are aligned, and the SHG fields thus constructively interfere, generating the maximum reflected intensity. On the other hand, destructive interfere will occur at $\Delta\theta = 60°$. Leveraging the partial analytical expression from ♯SHAARP.*ml*, we found that varying the $Al_2O_3$ thickness (500 ± 200 μm) can reduce the total refleted SHG intensity by 7-31% together with the absorption of $MoS_2$ at $2\omega$ frequency (using the imaginary dielectric constants of $MoS_2$ as $\varepsilon_I^{\omega} = 0$ and $\varepsilon_I^{2\omega} = 14.6$)[81], as shown by the red area. Detailed discussion on the substrate effect can be found in **Supplementary Note7, Figure. S7**. Our simulation results using ♯SHAARP.*ml* reproduced the sign change of $\kappa$ and indicate that the substrate effect may account for the scattered $\kappa$ values lying below the $\cos 3\Delta\theta$ curve as measured in the previous study.[79]

**Summary**

In summary, we have developed a comprehensive theoretical framework and implemented it into an open-source package (♯SHAARP.*ml*) for nonlinear optical analysis of multilayer systems including slabs and heterostructures, extending the existing capabilities of the prior ♯SHAARP.*si* package for single-interface systems. In addition to arbitrary materials properties such as symmetry, absorption, orientations, and dispersion, ♯SHAARP.*ml* also allows multiple reflections of both inhomogeneous and homogeneous waves at $\omega$ and $2\omega$ frequency, editable heterostructure schemes for versatile materials systems, integrated Maker fringes and polarimetry capabilities, and flexible probing conditions for both transmission and reflection geometries. The experimental and theoretical analyses based on various nonlinear optical crystals and multilayers help validate the capabilities and accuracy of ♯SHAARP.*ml* in the determination of nonlinear optical susceptibilities, crystal symmetries, and ferroelectric polarization directions. Seven material systems, namely α-quartz, α-quartz with Au coating, $LiNbO_3$, KTP, $ZnO//Pt//Al_2O_3$, $LiNbO_3$//α-quartz and twisted bilayer $MoS_2$ are chosen to benchmark ♯SHAARP.*ml* against our experimental measurements. The resulting absolute nonlinear optical susceptibilities and their relative ratios of all five cases show excellent agreement with the reported values. The successful demonstrations for the quartz+Au and $ZnO//Pt//Al_2O_3$ cases highlight the capabilities of modeling multiple reflection in a near Fabry-Perot condition. The simulation of a bilayer system with two SHG active media reveals the ability to accurately model SHG interference contrast imaging of otherwise undifferentiable ferroelectric domain states. The combined Maker fringes and SHG polarimetry capabilities of ♯SHAARP.*ml* make it a comprehensive analytical modeling tool for the optical metrology of new materials and heterostructures.

Looking forward, we expect that ♯SHAARP.*ml* can broadly streamline research in nonlinear optics. The complete and accurate analytical framework with editable assumptions from ♯SHAARP.*ml* can provide nonlinear optical solutions in an on-demand modality. As more integrated nonlinear optical devices and new topological superlattices are being developed, the capability of modeling these heterostructure can thus be an effective way to design, characterize, and optimize nonlinear optical response from complex systems. Furthermore, ♯SHAARP.*ml* provides a unique programmable platform for future extensions to new functionalities, such as other three-wave mixing processes, magnetic-dipole or quadrupole induced nonlinear optical effects, Gaussian beams with finite beam size, and inhomogeneous material systems.

## METHODS

### Sample Preparation

Both α-quartz and $LiNbO_3$ single crystals were obtained from MTI Corporation. The $(11\bar{2}0)$ and $(0001)$ oriented $LiNbO_3$, namely X-cut and Z-cut, were utilized in the analysis. Since the definition of X-cut $LiNbO_3$ from MTI is distinct from the orientations used in other analyses,[85,86] we have used the Miller indices for clarity. The X-cut and Y-cut KTP crystals were obtained from CASTECH Inc (Conex Systems Technology, Inc.). The $ZnO//Pt//Al_2O_3$ was prepared using RF magnetron sputtering, and the detailed growth procedure can be found in the earlier work.[43]

### Second-harmonic generation:

The second harmonic generation measurements were performed using a Ti: Sapphire femtosecond laser system with the central wavelength at 800 nm (1 kHz, 100 fs). The 1550 nm (1 kHz, 100 fs) was generated through an optical parametric amplifier, pumped by the 800 nm amplified laser. The SHG polarimetry measurements were performed using a combination of a zero-order half waveplate for the incident beam and an analyzer for the SHG signals. The polarization (azimuthal angle $\varphi$) of the incident linearly polarized light was rotated by the half-wave plate. The analyzer was set either parallel or perpendicular to PoI, equivalent to *p*- and *s*-polarized SHG, respectively. The polarized SHG was then filtered by the band pass filter to avoid additional spectrum contribution from the laser and samples. The Maker fringes measurements were performed by tilting samples while keeping incident and detecting polarization fixed. The rotation center of the sample stage is confirmed to be along the beam path to minimize the beam drift during the experiment. A photomultiplier tube (PMT) was used to collect SHG signals. The detected signals were further processed by the lock-in amplifier (SR830) to remove additional noise before feeding into the home-developed LabView program. The SHG fittings were then conducted using the expression generated by the ♯SHAARP. All the SHG coefficients from the literature are recalibrated using Miller's rule before the comparison.[87]

## Data Availability

The data that support the findings of this study are available from the corresponding authors upon reasonable request.

## Code Availability

The ♯SHAARP.*ml* is available through GitHub (https://github.com/bzw133/SHAARP.ml), and the documentation of the ♯SHAARP.*ml* can be accessed through ReadtheDocs (https://shaarpml.readthedocs.io/en/latest/).

## Acknowledgment

The ♯SHAARP open-source package is supported by the US Department of Energy, Office of Science, Basic Energy Sciences, under Award Number DE-SC0020145 as part of the Computational Materials Sciences Program. R.Z., B.W., A.S., L.-Q.C., and V.G. were supported by the U.S. Department of Energy, Office of Science, Basic Energy Sciences, under Award No. DE-SC0020145. J.H. acknowledges support from National Science Foundation under NSF DMR-2210933. L.W. was supported by NSF Research Experiences for Undergraduates (REU), DMR-1851987. Part of this work was performed under the auspices of the U.S. Department of Energy by Lawrence Livermore National Laboratory under Contract DE-AC52-07NA27344 (B.W.). R.Z. also received support from the NSF MRSEC Center for Nanoscale Science, DMR-2011839, for optical characterizations. R.Z. and V.G. acknowledge useful discussions with Dr. Zhiwen Liu. R.Z. acknowledges the analytical HH expressions from Sankalpa Hazra, the help with Au coating from Jeff Long and discussions with Dr. Yakun Yuan.

## Author Contribution

R.Z. and B.W. contributed equally to this work. R.Z., B.W., L.Q.C, and V.G. initiated the project. R.Z., B.W., L.Q.C, and V.G. developed the theory of SHG. R.Z., B.W., and L.W. contributed to code and GUI development. R.Z., B.W., and A.S. prepared the manual for ♯SHAARP.*ml*. R.Z. and J.H. collected experimental data. R.Z., B.W., L.Q.C., and V.G. carried out the SHG analysis. V.G. and L.Q.C. procured funding and supervised the project. R.Z., B.W., L.W., A.S., L.Q.C., and V.G. contributed to the manuscript preparation.

## Competing Interest

The authors declare no competing interests.

## Figure Legends

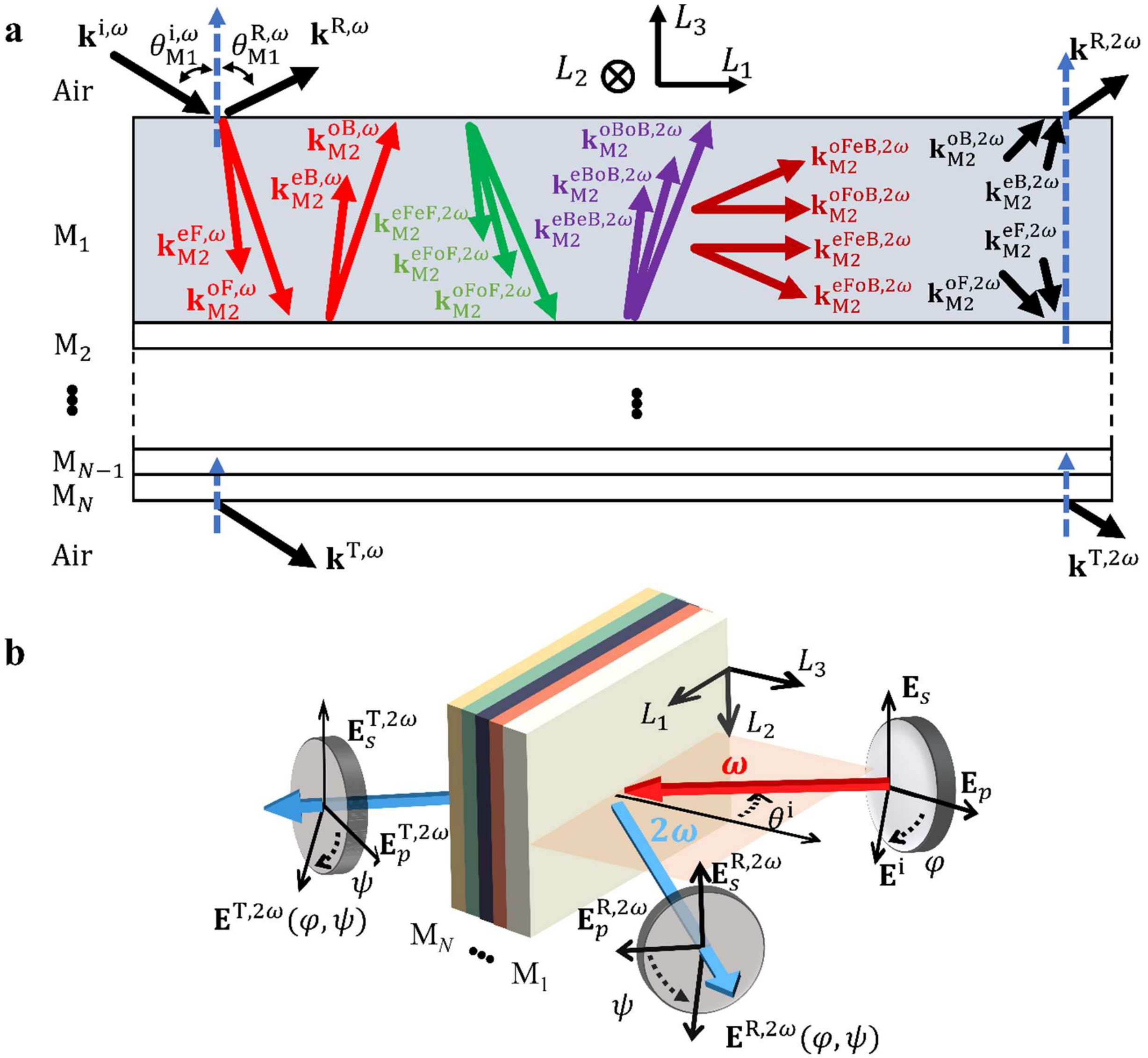

**Figure 1. Ray diagram and SHG measurement geometry. a.** The ray diagram of birefringent linear and nonlinear waves in the heterostructure. The $M_1$ layer is set to be SHG active. Both $\mathbf{k}^{\mathrm{eFeB},2\omega}$ and $\mathbf{k}^{\mathrm{oFoB},2\omega}$ are propagating parallel to layers. Different colors are used to distinguish different waves and are not indicative of their frequencies. **b** The SHG probing geometry. $(L_1, L_2, L_3)$ is the lab coordinate system. Red and blue rays are the fundamental beam at $\omega$ and SHG waves at $2\omega$, respectively. $\theta^{\mathrm{i}}$ is the angle of incidence, and the light red plane is the PoI, indicated by the $L_1 - L_3$ plane. The layers are subsequently labeled from $M_1$ to $M_N$.

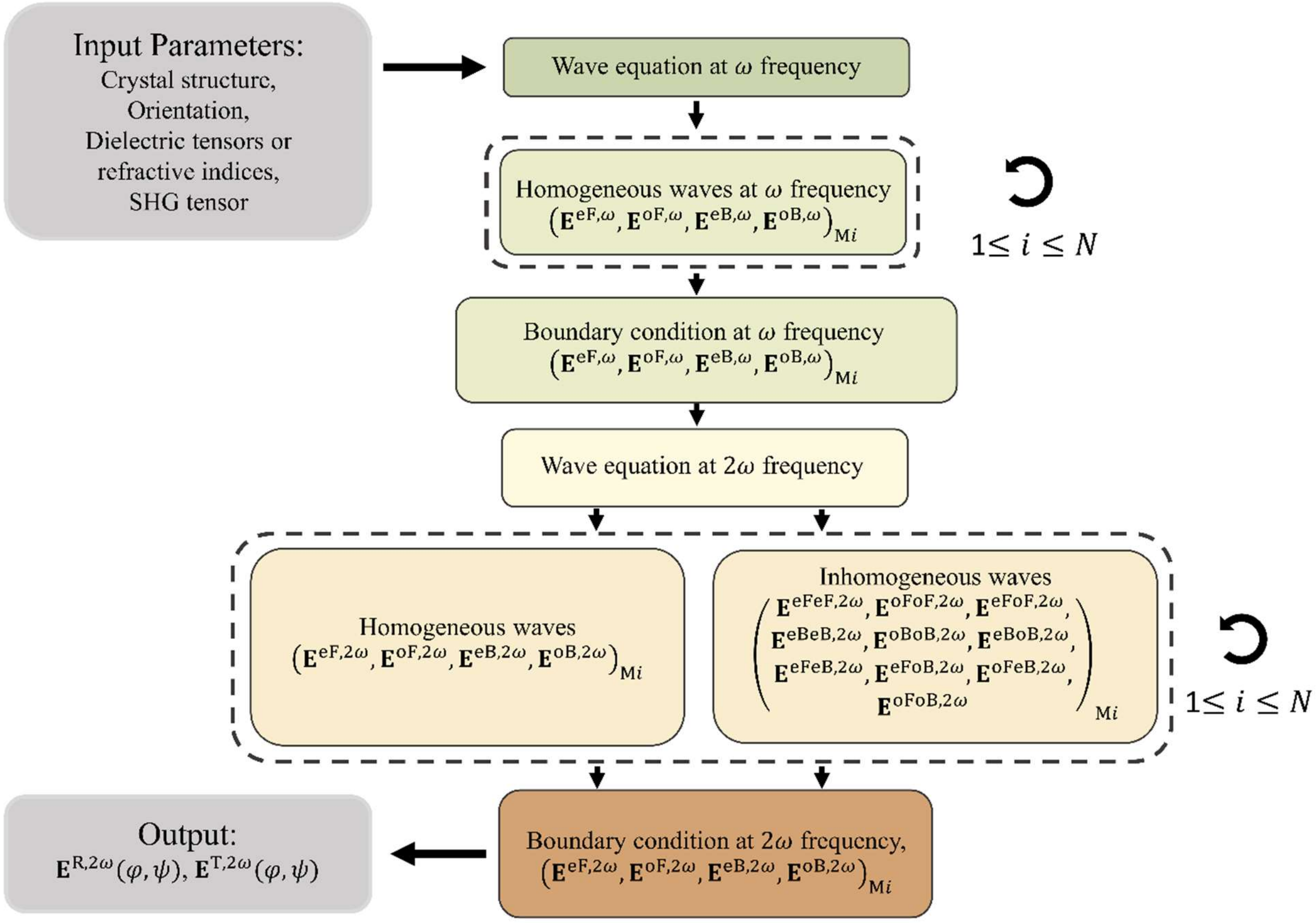

Figure 2. **Calculation procedures for ♯SHAARP.*ml*.** The dashed regions indicate repeating processes for all layers in the heterostructure.

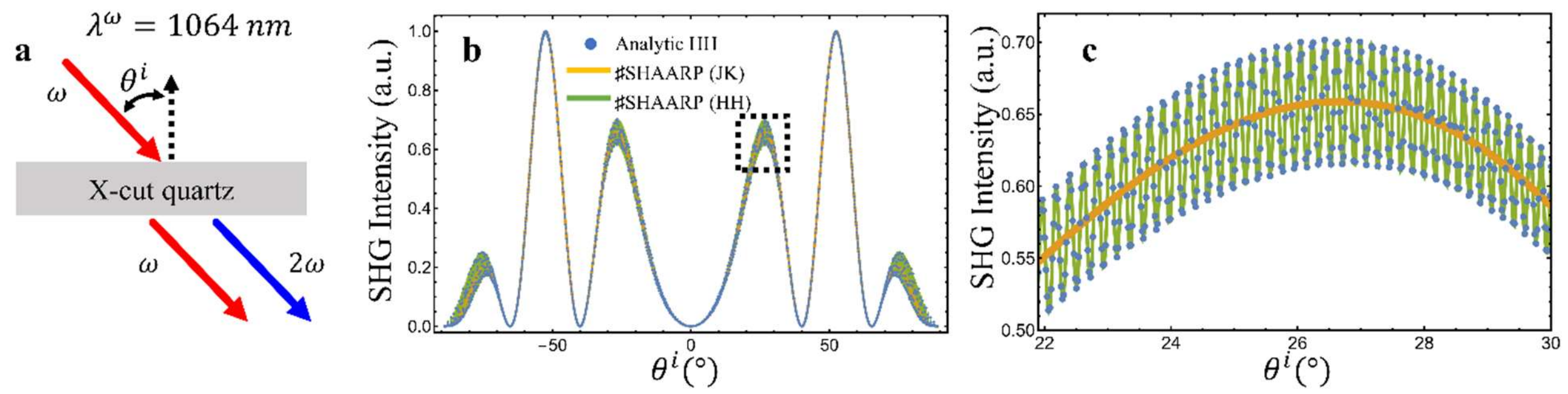

**Figure 3. Comparison of Maker fringes results between ♯SHAARP.*ml* and early analytical Herman & Hayden's and Jerphagnon & Kurtz's models.** (**a**) Schematic of Maker fringes condition using 300 µm X-cut quartz. The fundamental wavelength is 1064 nm. Red is fundamental light, and blue represents the generated SHG response. $\theta^i$ is the angle of incidence. Polarizations of both the fundamental and the SHG waves are set to be *p*- polarized. (**b**) SHG Maker fringes patterns obtained using Herman & Hayden's analytical expressions (analytic HH) and ♯SHAARP.*ml* analysis using Herman & Hayden's modeling condition, ♯SHAARP(HH), and Jerphagnon & Kurtz modeling condition, ♯SHAARP(JK). (**c**) Magnified region of (**b**) as indicated by the dashed box in (b).

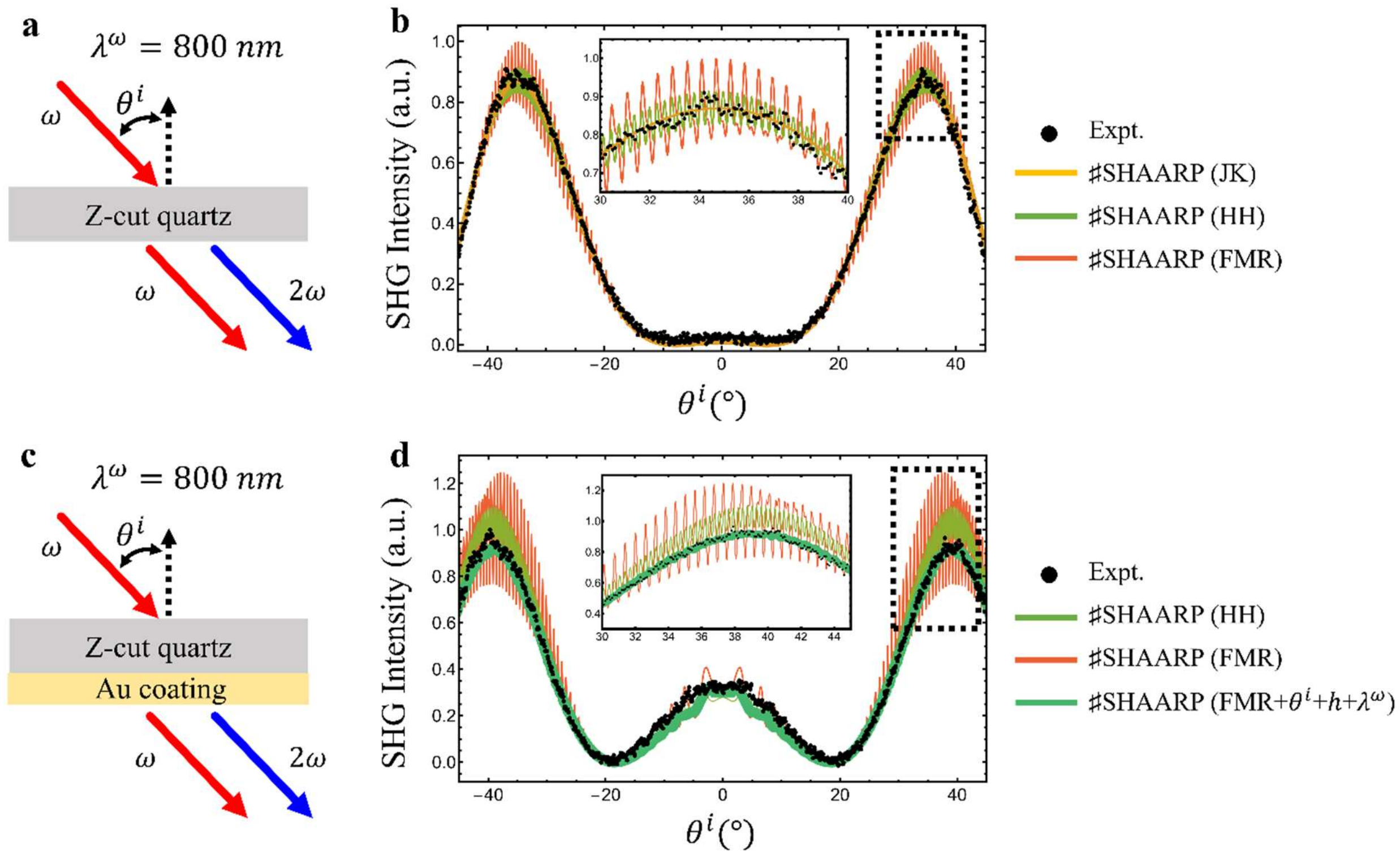

Figure 4. **Experimental verifications of ♯SHAARP.*ml* and influence of full multiple reflections using Maker fringes technique.** (**a**) Schematic of the experimental condition using Z-cut quartz. (**b**) The comparison among Maker fringes patterns from experiment and different modeling conditions based on the geometry in (a). The inset is the zoomed-in Maker fringes highlighted in the dashed area. (**c**) Schematic of the experimental condition using Z-cut quartz with a backside Au coating. (**d**) The comparison among Maker fringes patterns from experiment and different modeling conditions based on the geometry in (c). JK, HH, FMR and FMR+$\theta^i$+$h$+$\lambda^\omega$ represent JK method, HH method, full multiple reflections of linear and nonlinear waves, and averaged Maker fringes with a span of angles of incidence ($\theta^i$), crystal thicknesses ($h$) and wavelength of fundamental light ($\lambda^\omega$) due to a finite bandwidth based on ♯SHAARP(FMR). The fundamental wavelength $\lambda^\omega$ is 800 nm.

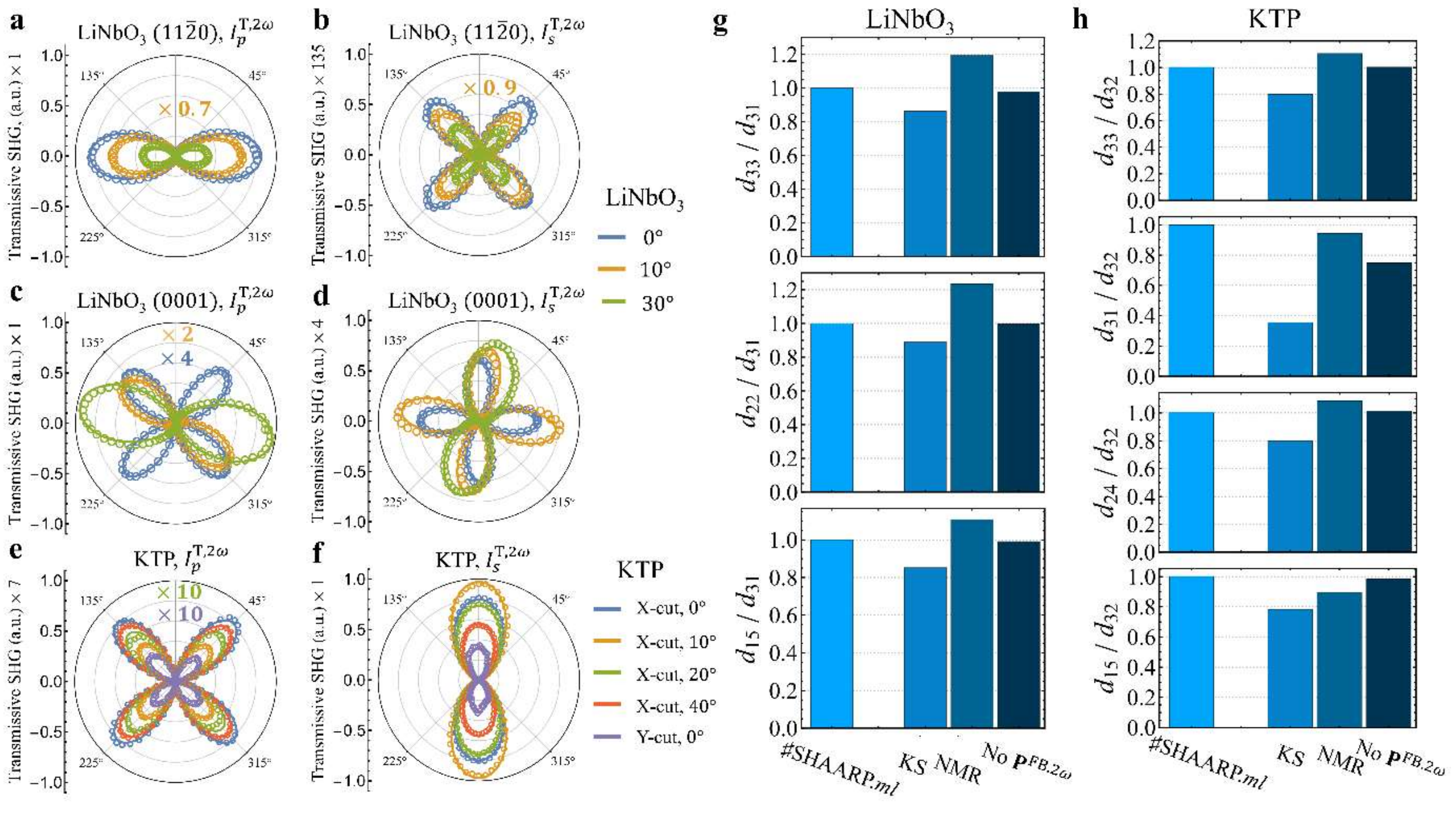

Figure 5. **Case studies of LiNbO3 and KTP single crystals at 1550 nm.** (**a**,**b**) The *p*- and *s*- polarized SHG response of $LiNbO_3$ (11$\bar{2}$0) as a function of incident polarization direction ($\varphi$). (**c**,**d**) The *p*- and *s*-polarized SHG response of $LiNbO_3$ (0001) as a function of incident polarization direction ($\varphi$). (**e**,**f**) The *p*- and *s*- polarized SHG response of KTP (100) and KTP (010) as a function of incident polarization direction ($\varphi$). (**g**,**h**) Comparison of extracted SHG coefficients ratios among different modeling conditions for (g) $LiNbO_3$ and (h) KTP. KS is Kleinman's symmetry. NMR stands for no multiple reflections of linear waves and nonlinear inhomogeneous waves. No $\mathbf{P}^{\mathrm{FB}}$ represents the case when the nonlinear polarizations generated by mixtures of forward and backward waves are ignored.

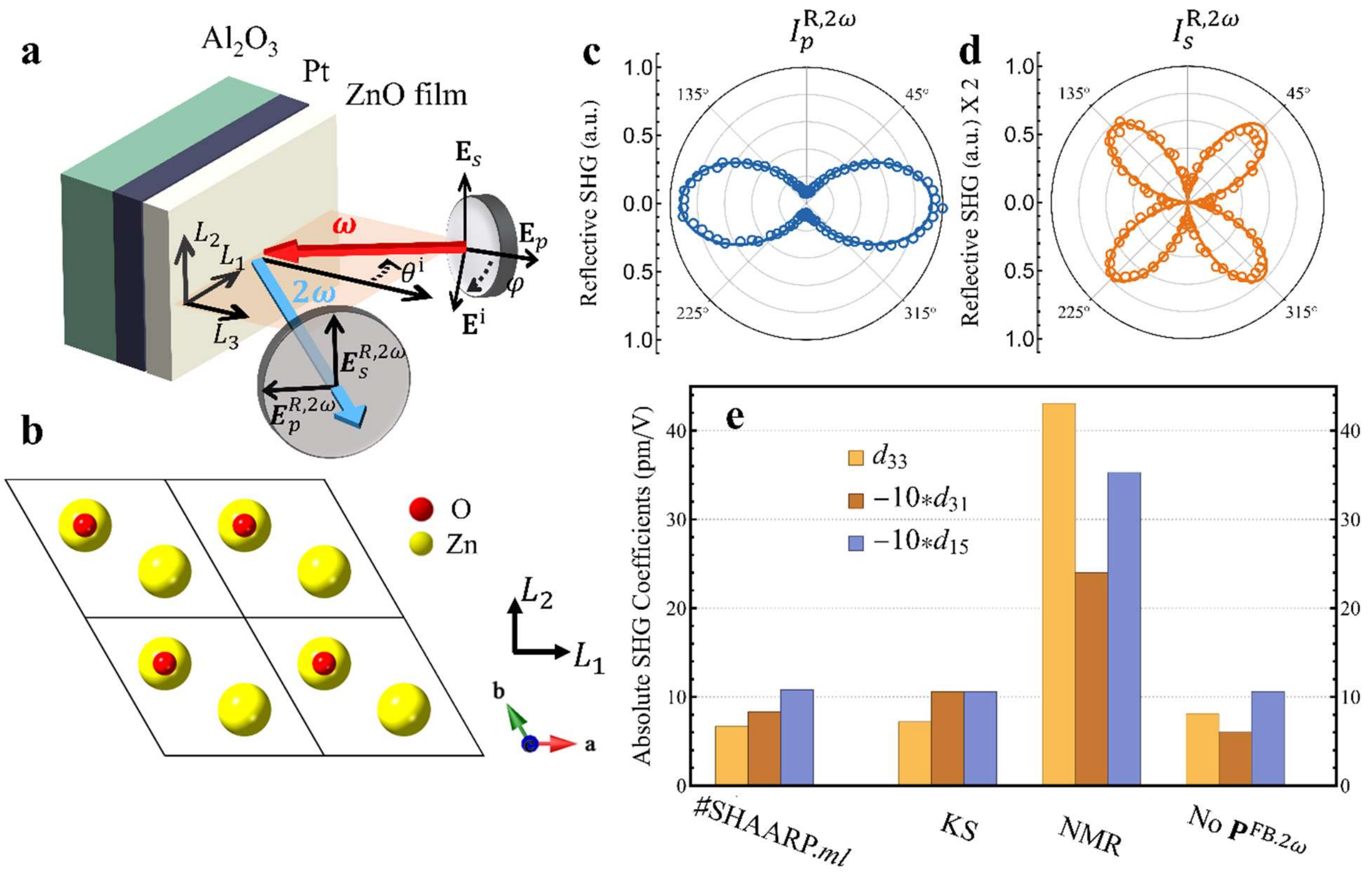

Figure 6. **Second harmonic generation analysis of ZnO//Pt//$Al_2O_3$ thin film at 1550 nm.** (**a**) The probing geometry of ZnO//Pt//$Al_2O_3$ heterostructure. The red beam is the fundamental ray, and the blue is the generated SHG response. The light red plane represents the plane of incidence parallel to $L_1 - L_3$ plane. The $p$- and $s$- polarized SHG response ($I_p^{R,2\omega}$ and $I_s^{R,2\omega}$) are collected as a function of incident polarization ($\varphi$). Superscript R indicates the reflected waves. (**b**) the relations between the crystallographic coordinate system $(a, b, c)$ and lab coordinate system $(L_1, L_2, L_3)$. (**c,d**) The SHG polarimetry results collected at $\theta^{\mathrm{i}} = 45°$ for (c) $p$- polarized SHG intensity $I_p^{R,2\omega}(\varphi)$ and (d) $s$- polarized SHG intensity $I_s^{R,2\omega}(\varphi)$. (**e**) The comparison of extracted complete absolute SHG tensor ($d_{33}$, $d_{31}$, $d_{15}$) among full analysis (♯SHAARP.*ml*) and various assumptions. KS is Kleinman's symmetry. NMR stands for no multiple reflections of linear waves and nonlinear inhomogeneous waves. No $\mathbf{P}^{\mathrm{FB}}$ represents the nonlinear polarizations generated by mixtures of forward and backward waves are ignored.

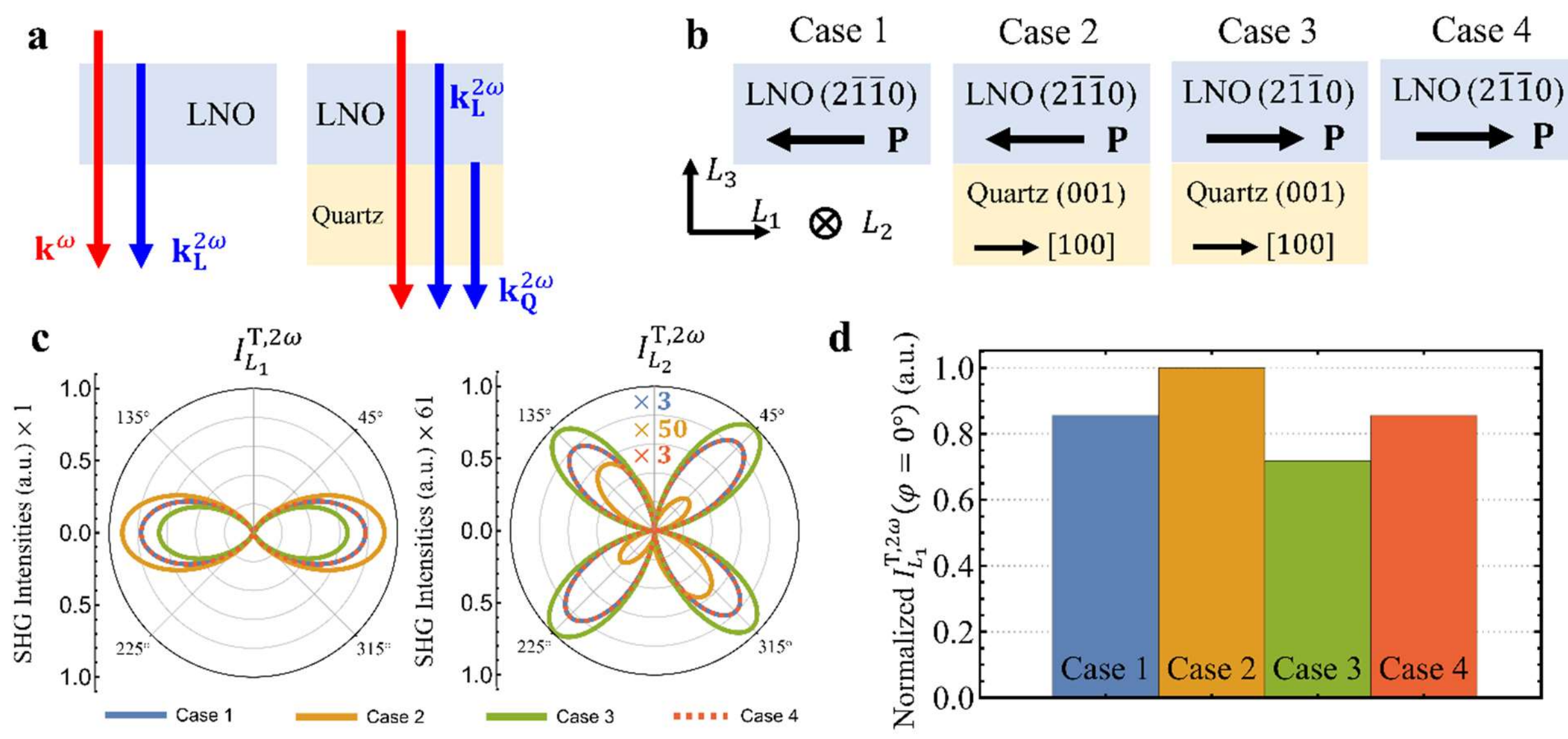

Figure 7. **Interferences of SHG intensities in LiNbO3 ( $2\bar{1}\bar{1}0$ ) and quartz ( 001 ) heterostructures.** (**a**) The ray diagrams of nonlinear waves in LiNbO3 (LNO) and LiNbO3//quartz. Red is fundamental light, and blue is SHG light. $\mathbf{k}_{\mathrm{L}}^{2\omega}$ and $\mathbf{k}_{\mathrm{Q}}^{2\omega}$ respectively refer to nonlinear waves generated by LiNbO3 and quartz. (**b**) Four cases used in the ♯SHAARP.*ml* simulation. The LiNbO3 ($2\bar{1}\bar{1}0$) and quartz (001) are used. The dark arrows in LiNbO3 indicate polarization directions parallel to the *c* axis. The dark arrows in quartz indicate the direction of [100]. Both case 2 and 3 use the same quartz for the interference study. (**c**) The resulting SHG polar plots for four cases in (b), subscripts $L_1$ and $L_2$ refer to SHG intensities polarized along $L_1$ and $L_2$ directions in (b). (**d**) The SHG intensity, $I_{L_1}^{T,2\omega}(\varphi = 0°)$, for four cases in (b).

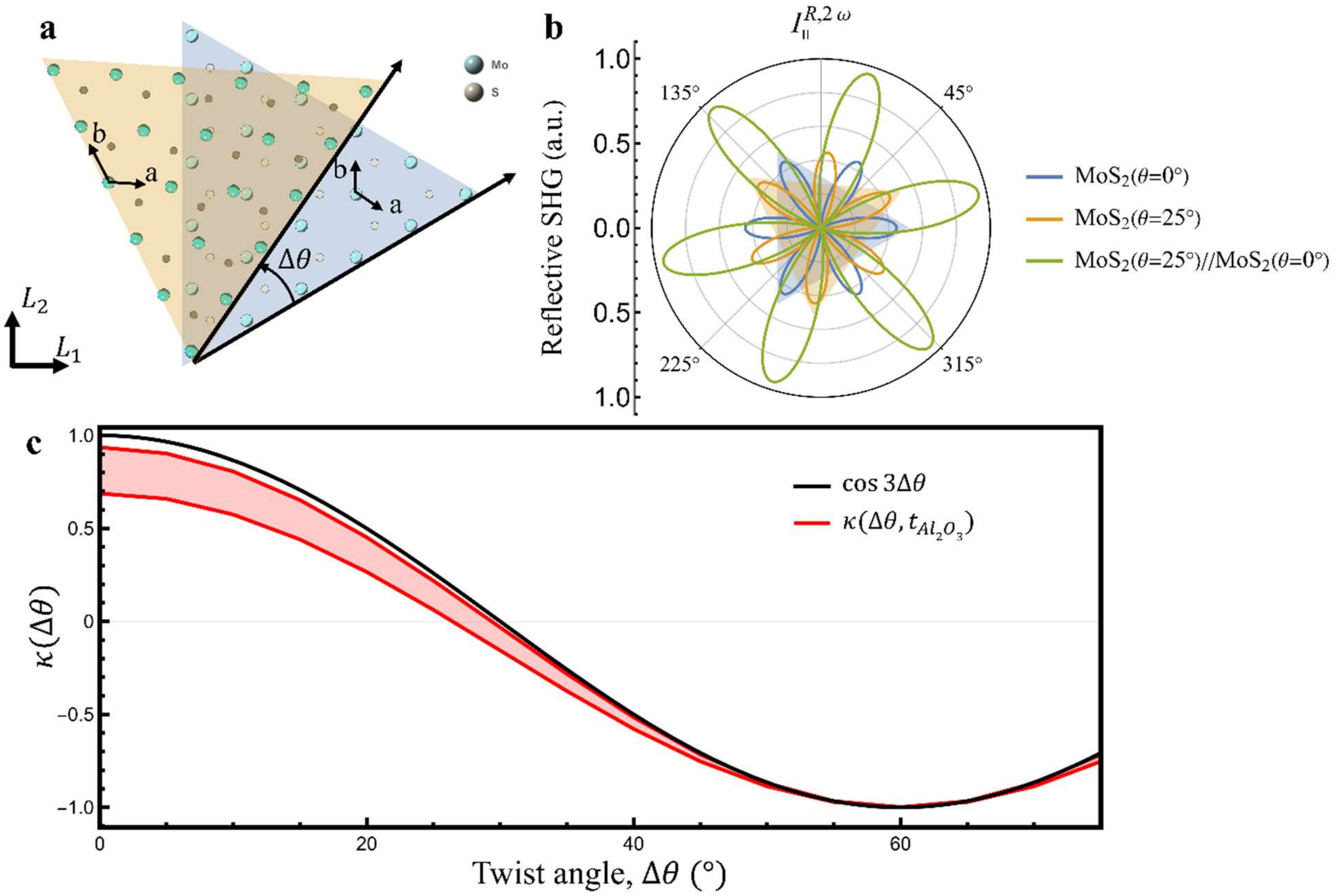

Figure 8. **SHG Polarimetry of twisted bilayer $MoS_2$.** (**a**) Crystal structure and relative orientations. $\Delta\theta$ represents the twist angle between the twisted top layer (orange) and the fixed bottom layer (blue). $L_1$-$L_2$ and $a$-$b$ represents LCS and CCS, respectively. (b) Resulting reflective SHG polarimetry ($I_{\parallel}^{R,2\omega}$) with parallel polarizer and analyzer conditions, where blue and orange represent monolayers and green stands for the bilayer. 0° in the polar plot refers to the electric field parallel to $L_1$ direction. The triangles indicate the orientations of monolayer $MoS_2$ in panel (a). (c) $\kappa$ as a function of the twist angle ($\Delta\theta$). The red area shows the variation of $\kappa$ due to absorption of $MoS_2$ and thickness variation of $Al_2O_3$ substrate.

**Supplementary Information for**

# Optical Second Harmonic Generation in Anisotropic Multilayers with Complete Multireflection of Linear and Nonlinear Waves using ♯SHAARP.*ml* Package

## Supplementary Note 1:

**Figure S1** summarizes the differences in the various modeling methods, including JK, HH, and ♯SHAARP.*ml*. In the JK method, the transmitted light at $\omega$ at the top surface is used to calculate the nonlinear polarizations (inhomogeneous wave at $2\omega$, abbreviated as Inhomo). The boundary conditions at $2\omega$ are carried out separately at each interface (as indicated in each dashed region), and their corresponding waves at each interface are listed in Figure S1. Notably, the top surface only includes forward waves, but the bottom surface contains backward homogeneous (abbreviated as Homo), as labeled as $\mathbf{k}^{\mathrm{B},2\omega}$. This means both homogeneous wave and inhomogeneous wave at $2\omega$ will only propagate once, and back reflected waves will not reach the top interface. On the other hand, the HH method performs the same treatment of linear waves as the JK method. The key differences are: (1) the boundary conditions at $2\omega$ at all interfaces are solved simultaneously, as highlighted by the dashed region in the HH method, enclosing Inhomo and Homo waves and all interfaces; (2) The Homo waves contain both $\mathbf{k}^{\mathrm{F},2\omega}$ and $\mathbf{k}^{\mathrm{B},2\omega}$ at all interfaces suggesting multiple reflections of homogeneous waves at $2\omega$ are considered. Furthermore, in ♯SHAARP.*ml*, both $\mathbf{k}^{\mathrm{F},\omega}$ and $\mathbf{k}^{\mathrm{B},\omega}$ at all interfaces are considered, and boundary conditions at all interfaces at $\omega$ are solved simultaneously. This means multiple reflections at linear frequency are involved, as shown in the left dashed region in ♯SHAARP.*ml*. The resulting $\mathbf{k}^{\mathrm{F},\omega}$ and $\mathbf{k}^{\mathrm{B},\omega}$ are used in determining all ten nonlinear polarizations in each SHG active medium. Solving the boundary conditions at all interfaces at $2\omega$, containing all forward and backward waves of both Inhomo and homo will lead to the full consideration of multiple reflections at both $\omega$ and $2\omega$ frequencies.

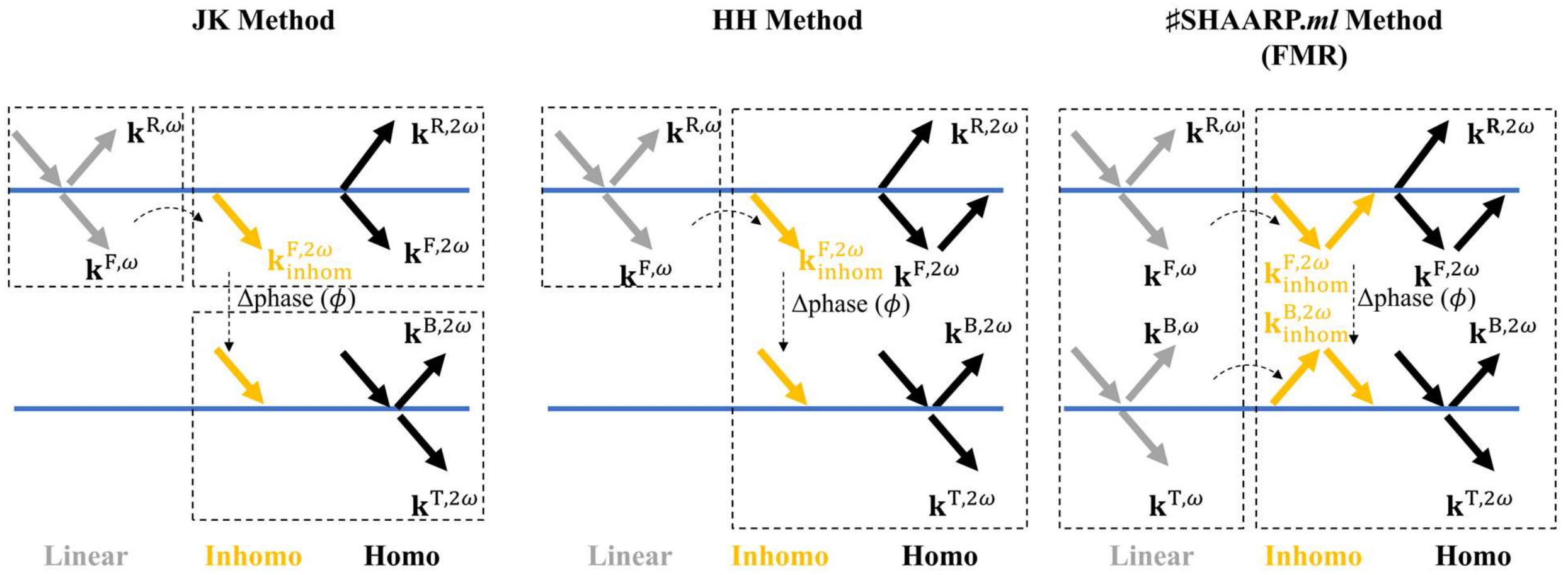

Figure S1. The differences in the nonlinear optical modelings among JK, HH, and ♯SHAARP.*ml* methods. The dashed regions represent independent boundary conditions. Linear, Inhomo, and Homo represent linear waves at $\omega$, inhomogeneous waves at $2\omega$, and homogeneous waves at $2\omega$, respectively.

**Supplementary note 2:**

The influences of averaging the angle of incidence and thickness on the Maker fringes patterns are investigated, as shown in **Figure S2**. Experimentally, the lens with a focusing distance of 10 cm (f=10 cm) is used, and the fundamental beam has a diameter of 5 mm (d≈ 5 mm). Thus, the convergence angle is estimated to be $Tan^{-1}(\frac{d}{f}) \approx 3°$. The spot size focused on the sample at $\lambda^{\omega} = 800$ nm is around 50 µm, and a near 10 µm thickness variation is observed across 10 mm × 10 mm sample. Thus, the thickness variation within the spot size is estimated to be 50 nm. Thus, averaged Maker fringes patterns with a 3° binning widow for $\theta^{i}$ and a 50 nm binning window for sample thickness ($h$) are investigated. The bandwidth of 800 nm laser was measured to have $\pm 5$ nm variation. The variation in the wavelength ($\lambda$) has a similar effect as the thickness variation, and both effects contribute to the phase accumulated throughout the crystal, i.e. $\phi = \frac{2\pi}{\lambda} h$. Thus, the variation in $\lambda$ is estimated to result in ~0.6% variation in phase, equivalent to ~0.7 μm variation in $h$ for a 120 μm thick crystal.

By averaging $\theta^i$, the ♯SHAARP(FMR) can be effectively smoothed and the resulting ♯SHAARP(FMR+$\theta^i$) agrees well with the experimental observation. Averaging $h$ based on ♯SHAARP(FMR+ $\theta^i$ ), the obtained ♯SHAARP(FMR+ $\theta^i$ +$h$) overlaps with ♯SHAARP(FMR+$\theta^i$). Further averaging $\lambda$ based on ♯SHAARP(FMR+$\theta^i$+$h$), the obtained ♯SHAARP(FMR+$\theta^i$+$h$+$\lambda$) can provide better agreement with experimental results especially at low angles of incidence. Thus, this averaging study suggests the spread of angles of incidence, thickness variation of the crystal and finite bandwidth of the fundamental wavelength in the experiment can contribute to smearing out oscillations calculated from ♯SHAARP(FMR).

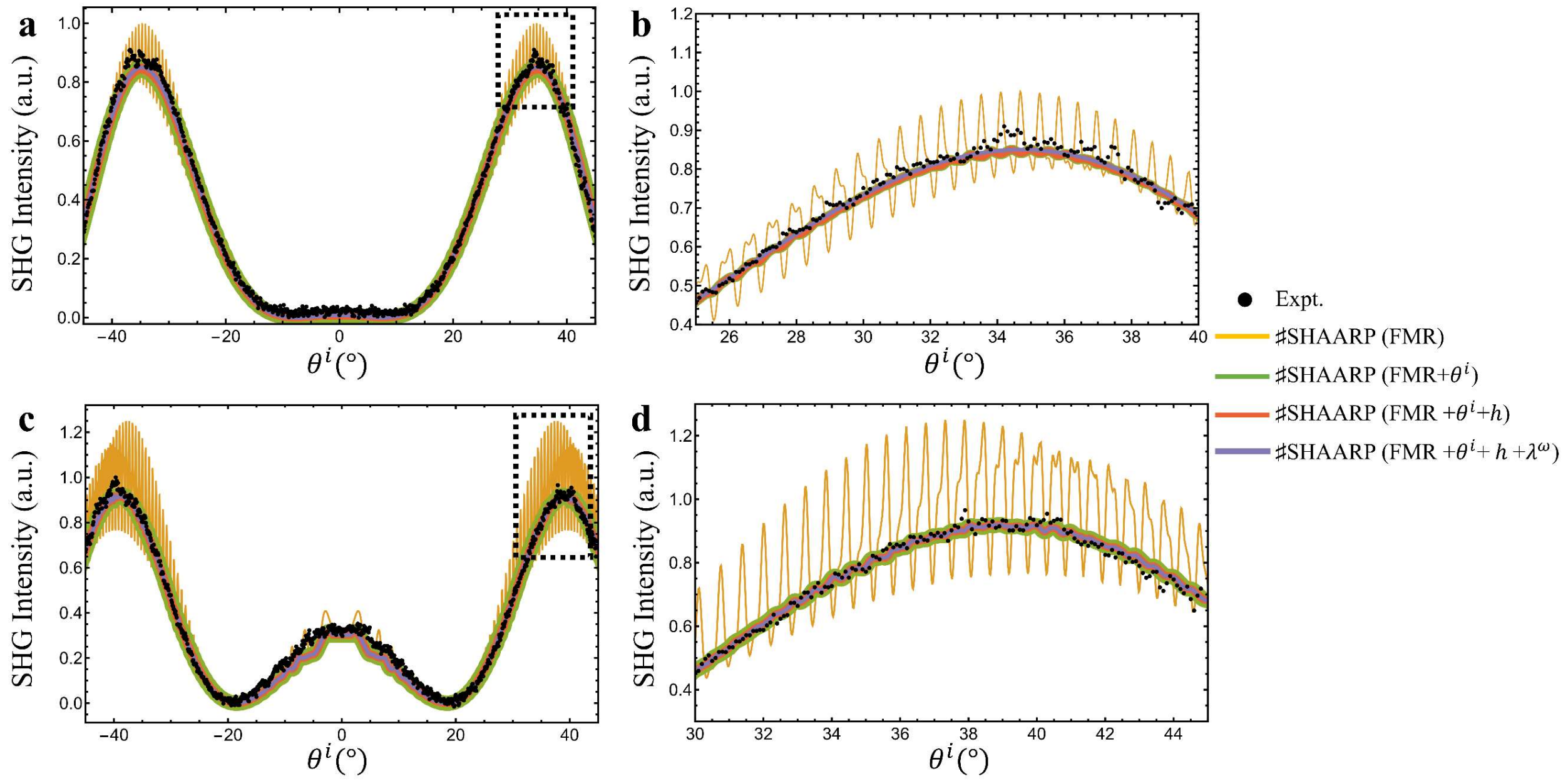

Figure S2. **The influence of averaging the angle of incidence ($\theta^i$), averaging sample thickness ($h$) and averaging wavelength ($\lambda$) on the Maker fringes patterns.** (a,b) Maker fringes patterns using Z-cut quartz. (c,d) Maker fringes patterns using Z-cut quartz with backside Au coating. (b) and (d) are zoomed-in regions indicated by the dashed area in (a) and (c), respectively. The black dots are experimental results (Expt.). ♯SHAARP(FMR) represents the full multiple reflections analysis. ♯SHAARP(FMR+ $\theta^i$ ) indicates averaging the angle of incidence based on ♯SHAARP(FMR). ♯SHAARP(FMR+$\theta^i$+$h$) stands for additional averaging thickness analysis based on ♯SHAARP(FMR+$\theta^i$). ♯SHAARP(FMR+$\theta^i$+$h$+$\lambda$) indicates further averaging wavelength analysis based on ♯SHAARP(FMR+$\theta^i$+$h$).

**Supplementary note 3:**

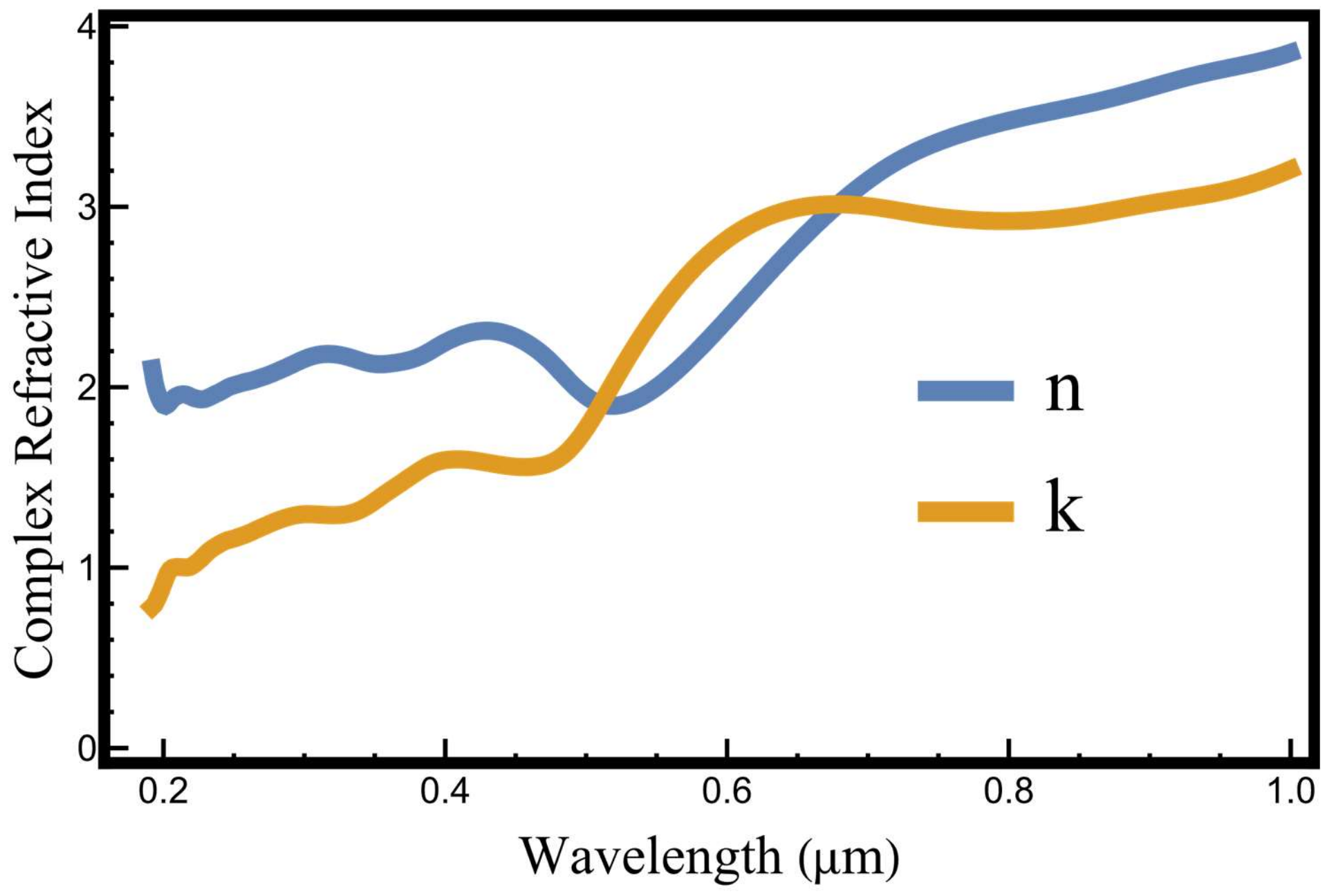

Figure S3. **The complex refractive index of the Au layer determined using spectroscopic ellipsometry.** The resulting layer thickness is determined to be 13.9 nm.

**Supplementary note 4:**

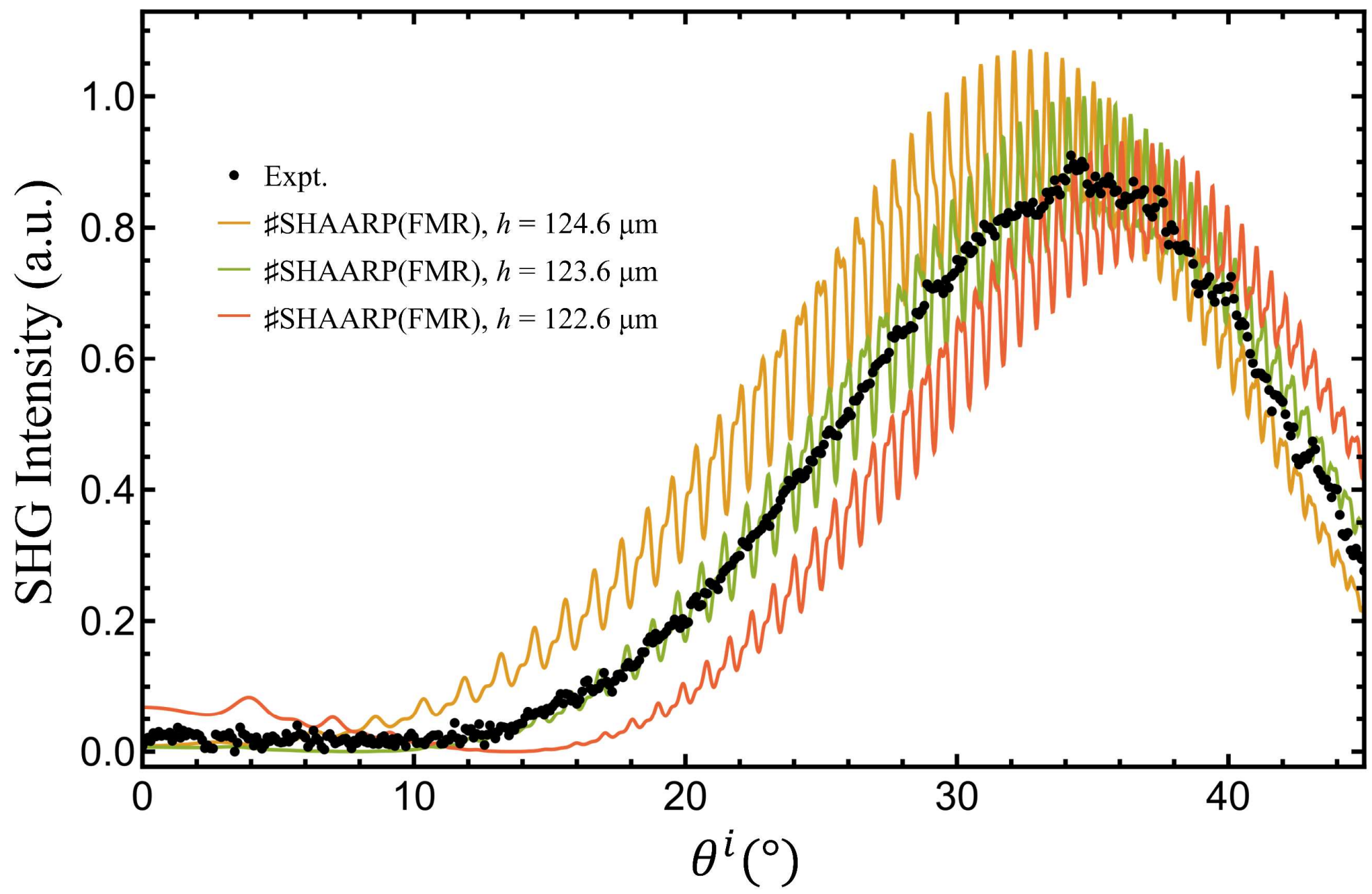

Figure S4. **The Maker fringes pattern of Quartz with different thicknesses.** The dot corresponds to the experimental results shown in **Figure 4b**. The $h$ is the crystal thickness used in the simulation. The thickness of the crystal near the probing area is determined to be 123.6 $\mu m$, while $\pm 1$ $\mu m$ in h leads to a large change in the Maker fringes pattern.

**Supplementary Note 5:**

The phase calculation methods play a critical role in the predicted Maker fringes pattern. **Figure S5** compares two different phase calculation methods and explores the influence on the obtained Maker fringes pattern. Conventionally, the phase of electromagnetic waves accumulated throughout the slab is calculated using $\phi = h \cdot \mathbf{k} \cdot (0,0,-1)$, where $\phi$, h and $\mathbf{k}$ are the accumulated phase, sample thickness and wave vector. Here, only the wave vector along $L_3$ direction is considered for the wave accumulation, and we denote this case as the vertical phase (VP). Additionally, we have performed the full phase analysis of the accumulated phase (FP). Mathematically, FP case uses $\phi = h \cdot \mathbf{k} \cdot (\tan\theta, 0, -1)$, where $\theta$ is the refractive angle. In FP, both tangential and vertical phases are considered. However, as demonstrated in **Figure S5a** and **Figure S5b** for pure quartz, and **Figure S5c** and **Figure S5d** for quartz+Au, the FP case (represented as ♯SHAARP(FMR+FP)) deviates from the experimental results significantly, suggesting taking tangential phase can cause large errors. Such discrepancy may come from the fact that a finite beam is used in the experiment, where the beam overlap is essential in the experiment and SHG analysis.

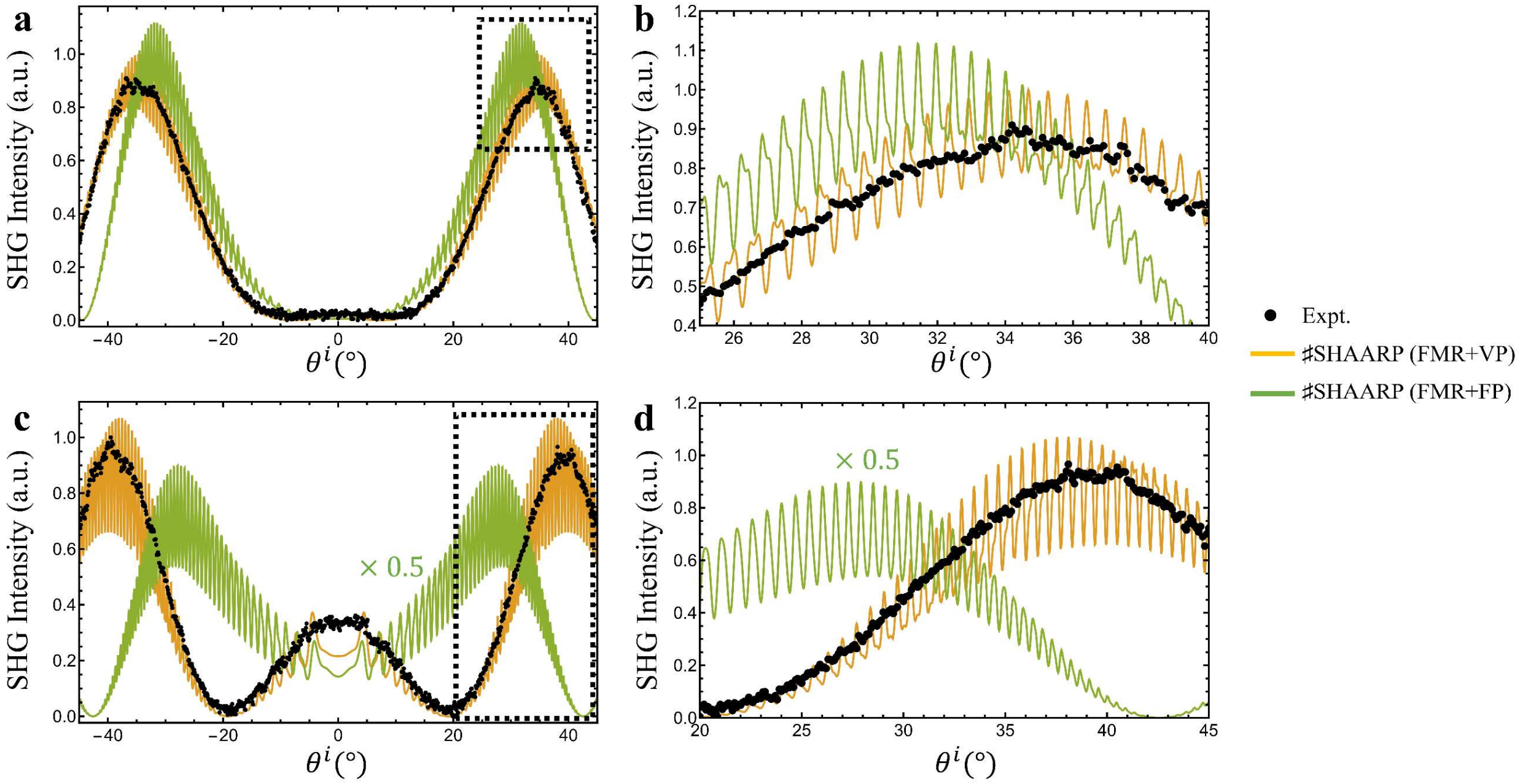

Figure S5. **The influence of different phase calculation methods on the Maker fringes patterns.** (a,b) Maker fringes patterns using Z-cut quartz. (c,d) Maker fringes patterns using Z-cut quartz with backside Au coating. (b) and (d) are zoomed-in regions indicated by the dashed area in (a) and (c), respectively. The black dots are experimental results (Expt.). FMR represents the full multiple reflections analysis. ♯SHAARP(FMR+VP) indicates the full multiple reflections with the vertical phase of waves used in the analysis. ♯SHAARP(FMR+FP) represents the full multiple reflections with the full phase of waves used in the analysis.

## Supplementary Note 6:

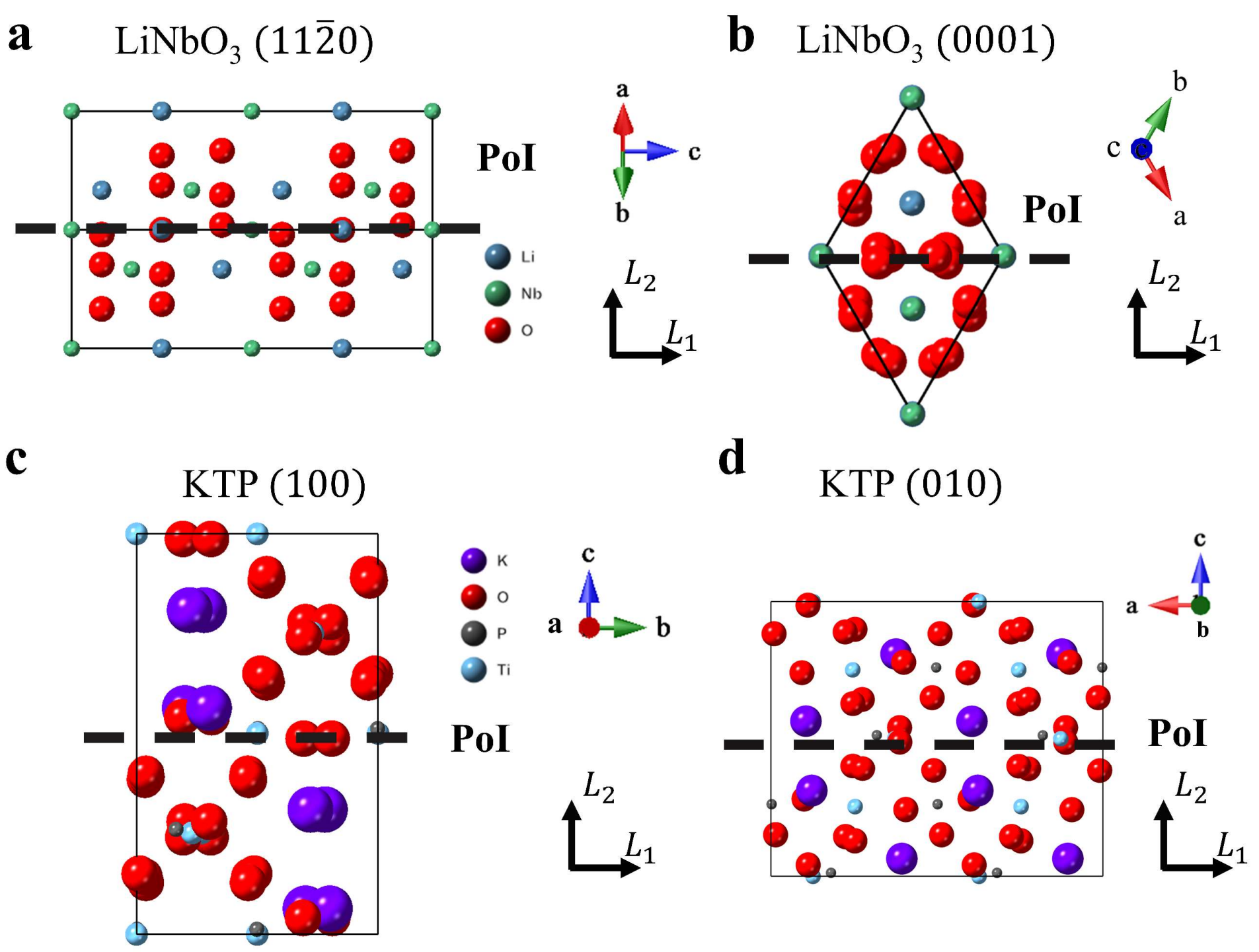

Figure S6. Relations between crystal orientation (CCS) and lab coordinate system (LCS) for $LiNbO_3$ and KTP. The $(L_1, L_2, L_3,)$ is LCS, $(a, b, c)$ is CCS. Dashed lines are the planes of incidence parallel to the $L_1 - L_3$ plane.

**Supplementary Note 7:**

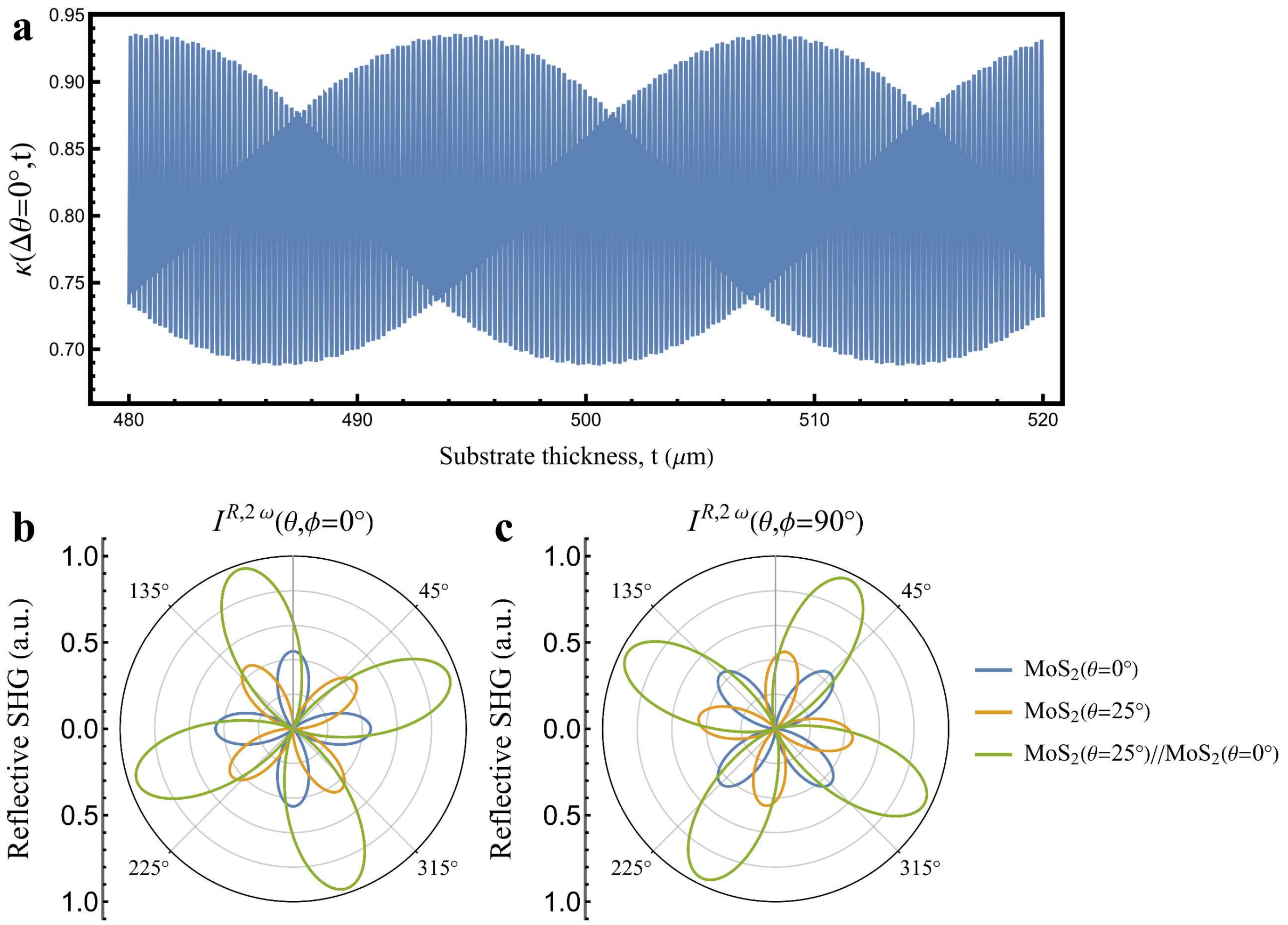

Figure S7. Thickness dependent $\kappa$ in RA geometry and polarimetry with fixed analyzer (FA geometry) of twist bilayer $MoS_2$. (a) $\kappa$ function of bilayer $MoS_2$ with a twist angle $\Delta\theta = 0°$ in RA configuration. The $\kappa$ function for the thickness range 500 ± 200 μm follows the same periodicity as shown in (a). (b,c) SHG polarimetry with a fixed analyzer set at 0° and 90°, respectively. The azimuthal angle at $\varphi = 0°$ refers to the incident electric field along $L_1$ direction. The green curve shows the bilayer case with a twist angle $\Delta\theta = 25°$.